\documentclass[aps,pre,twocolumn,eqsecnum,superscriptaddress]{revtex4}    
\usepackage{amsmath}    % need for subequations
\usepackage{amssymb}
\usepackage{subfigure}
\usepackage{epsfig}
\usepackage{array}      % for controlling table column width

\DeclareMathOperator{\sech}{sech}

\begin{document}

%\title{Non-equilibrium thermodynamics of comminution in sheared granular materials}
\title{Grain fragmentation in sheared granular flow: weakening effects, energy dissipation, and strain localization}

\author{Charles K. C. Lieou}
\affiliation{Department of Physics, University of California, Santa Barbara, CA 93106, USA}
\author{Ahmed E. Elbanna}
\affiliation{Department of Civil and Environmental Engineering, University of Illinois, Urbana-Champaign, IL 61801, USA}
%\author{James S. Langer}
%\affiliation{Department of Physics, University of California, Santa Barbara, CA 93106, USA}
\author{Jean M. Carlson}
\affiliation{Department of Physics, University of California, Santa Barbara, CA 93106, USA}
\date{\today}

\begin{abstract}
We describe the shear flow of a disordered granular material in the presence of grain fracture using the shear-transformation-zone (STZ) theory of amorphous plasticity adapted to systems with a hard-core inter-particle interaction. To this end, we develop the equations of motion for this system within a statistical-thermodynamic framework analogous to that used in the analysis of molecular glasses. For hard-core systems, the amount of internal, configurational disorder is characterized by the compactivity $X = \partial V / \partial S_C$, where $V$ and $S_C$ are respectively the volume and configurational entropy. Grain breakage is described by a constitutive equation for the temporal evolution of a characteristic grain size $a$, based on fracture mechanics. We show that grain breakage is a weakening mechanism, significantly lowering the flow stress at large strain rates, if the material is rate-strengthening in character. We show in addition that if the granular material is sufficiently aged, spatial inhomogeneity in configurational disorder results in strain localization. We also show that grain splitting contributes significantly to comminution at small shear strains, while grain abrasion becomes dominant at large shear displacements.
\end{abstract}

\maketitle

\section{Introduction}

\begin{figure}[here]
\centering 
\includegraphics[width=.45\textwidth]{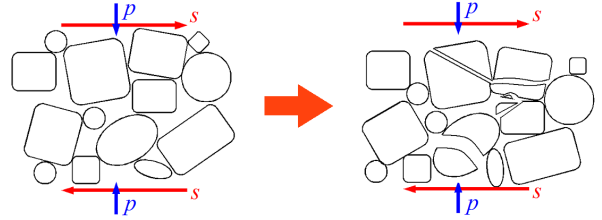} \caption{\label{fig:shear_fragmentation}(Color online) Schematic illustration of grain fragmentation under shear flow. The granular material is subject to a shear stress $s$ and a confining pressure $p$. Grains can undergo rearrangement or comminution upon the onset of shear stress.} 
\end{figure}

In this paper, we develop a statistical-thermodyanmic theory for the shear flow of a disordered granular material that undergoes grain breakage, depicted in Fig.~\ref{fig:shear_fragmentation}. Such a theoretical description is of fundamental importance in understanding practical problems involving granular systems, such as the rupture dynamics of a sheared fault gouge and the nucleation of earthquakes.

In doing so, we build upon our recent work~\cite{lieou_2012} in which we combine the basic elements of the shear-transformation-zone (STZ) theory of amorphous molecular plasticity \cite{falk_1998,langer_2011,manning_2007a,langer_2008,langer_2011a,langer_2011b,langer_2012} with Edwards' statistical theory of granular materials \cite{edwards_1989a,edwards_1989b,edwards_1989c,edwards_1990a,edwards_1990b} to construct a theory of shear flow in a noncrystalline system of thermalized hard spheres. The STZ theory is based on the premise that irreversible particle rearrangements occur at isolated flow defects (i.e.~STZ's); in deforming systems, STZ's are activated fluctuations which appear and disappear in response to thermal or mechanical noise. The plastic strain rate is proportional to the STZ density. In~\cite{lieou_2012}, we found that the density of STZ's is given by a Boltzmann-like factor of the form $\exp ( - v_Z / X)$. Here, $v_Z$ denotes the excess volume per STZ; $X$ is the compactivity, a measure of the amount of configurational disorder in the system, defined in terms of the extensive volume $V$ and the configurational entropy $S_C$ as
\begin{equation}\label{eq:compactivity_def}
 X = \dfrac{\partial V}{\partial S_C}.
\end{equation}
With a judicious choice of several parameters, the STZ theory has enabled us to interpret numerical simulations of a driven hard-sphere system~\cite{liu_2011,haxton_2012}, and to obatin useful insights about internal rate factors and the relations between jamming and glass transitions.

%%%%%%%%%%%%% FIGURE 1 %%%%%%%%%%%%
%%%%%%%%%%%%%%%%%%%%%%%%%%%%%%%%%%%%% 

The hard-sphere system is a simplified model of a granular material, composed of grains with no internal structure. Real granular systems, however, consists of grains that come in a variety of sizes and shapes. Often, these grains can break apart upon the application of a sufficiently large force, as seen in many simulations and geophysical field observations \cite{mair_2011,sammis_1987,reches_2010,han_2011,mair_2002,guo_2006,mair_1999,rice_2006, chambon_2006}. A theoretical description of grain fragmentation in sheared granular flow should provide plausible explanations to key features of those observations and offer predictive capabilities; such a theory is clearly of practical utility to the granular physics community at large.

Our goal, therefore, is to bridge the gap between macroscopic constitutive laws and microscopic physics of grain fragmentation, by drawing on the familiar concept of disorder from nonequilibrium statistical mechanics. Specifically, we shall propose an evolution equation that quantifies the breakage process, and combine that with the STZ theory of dense, sheared granular flow~\cite{lieou_2012}. While the compactivity $X$ as defined in Eq.~\eqref{eq:compactivity_def} remains a valid measure of the amount of configurational disorder, it is clear that new theoretical elements are needed in order to accurately describe these additional phenomena within our statistical-thermodynamic framework, and to understand the interplay between grain rearrangement and fragmentation processes. This is the purpose of our present paper.

The rest of the paper is structured as follows. Section II is devoted to a review of experimental findings and modeling efforts in the literature; the purpose is to highlight the aspects of observations in experiments and simulations, as well as to examine elements of existing theories, that motivate our theoretical model. We begin our main theoretical development in Sec.~III, where we review the first and second laws of thermodynamics for a driven granular system that undergoes grain breakage. In Sec.~IV we derive the STZ equations of motion. Then, in Sec.~V we deduce an equation of motion that governs the temporal evolution of the characteristic grain size $a$. We analyze the thermodynamics of heat dissipation and derive constraints on rate factors and quasistationary variables in Sec.~VI, and conclude our theoretical development with an equation of motion for the compactivity $X$ in Sec.~VII. In Sec.~VIII, we elucidate some features of the temporal behavior of the shear stress and the characteristic grain size, and discuss the implications of grain size reduction on the energy partition into fracture energy, plastic work and heat dissipation. In Sec.~IX we expand our development to describe shear localization, by introducing spatial heterogeneity in the initial conditions, as we did for amorphous, molecular systems~\cite{manning_2007a,manning_2009,daub_2009}. We shall show several different kinds of localization behaviors that can arise, and examine the implication of grain breakage on strain localization. Section X concludes our paper with a summary of our results and a discussion of future directions.

\section{Survey of experiments, simulations and theories of grain fragmentation in sheared granular flow}

Among the many simulations, experiments, and geophysical observations on grain breakage in the literature, Mair and Abe~\cite{mair_2011} modeled grains as aggregates of particles stuck together with elastic bonds, and used three-dimensional discrete element simulations to study the fragmentation processes in fault gouge; they found that grain splitting dominates under high normal stress and at small shear strains, while grain abrasion dominates under low normal stress and at large shear strains. Sammis and coworkers~\cite{sammis_1987} examined samples of exhumed fault gouge and concluded that the particle size distribution evolves into a self-similar distribution under generic conditions. These findings indicate that the confining pressure plays an important role in determining the occurrence of grain breakage and the size distribution of particles thus formed.

Reches and Lockner~\cite{reches_2010} and Han et al.~\cite{han_2011} found that extremely fine gouge particles lubricate the fault, and suggested that grain breakage is a weakening mechanism at seismic slip rates. Reches and Lockner additionally found significant strengthening at intermediate slip rates, but they concluded that this strengthening could be traced to factors such as dehydration due to heating as well as agglomeration. On the other hand, Mair et al.~\cite{mair_2002} found that grain breakage is a strengthening mechanism for a granular material composed of initially smooth, spherical grains but is neither intrinsically strengthening nor weakening for rough, angular grains; they attributed the strengthening seen in smooth grains to the increasing angularity as grains break apart. Guo and Morgan~\cite{guo_2006} performed distinct element simulations of grains constructed from clusters of circular particles, and found that grain fragmentation could be strengthening or weakening depending on the effect of fragmentation of grain angularity and elongation. These observations suggest that it is prudent to decouple the effects of grain fragmentation from the strengthening or weakening effects due to particle shape, roughness and friction in a basic theoretical model; this is the approach that we shall adopt in this paper.

Many others (e.g.~\cite{mair_1999,rice_2006,chambon_2006}) have identified the presence of localization in a sheared granular system; the particles in the shear band are often a few orders of magnitude smaller than particles outside the shear zone. These observations point to a feedback between plastic deformation and grain fragmentation within the shear zone; an attempt to elucidate this feedback mechanism, as well as its implications on shear strength, is of fundamental significance.

There have been few theoretical attempts to model grain breakage and track the temporal evolution of properties characterizing the state of the system. Among such attempts, the most notable is the work by Einav and coworkers~\cite{einav_2007a,einav_2007b,einav_2009}, who proposed a constitutive model based on the theory of breakage mechanics. They introduced an internal ``breakage'' variable to characterize the distance of the instantaneous particle size distribution from the ultimate size distribution, and postulated constitutive relations between that quantity and associated thermodynamic potentials within the framework of continuum mechanics. In this manner, they connected the energetics and micromechanics of grain comminution. The breakage mechanics model was used to compute the stress-strain behavior and quantify the energy budget in sheared fault gouge, and was able to account for permeability reduction in rocks. Lyakhovsky and Ben-Zion~\cite{lyakhovsky_2013} further refined the model to describe brittle instability and shear localization as a phase transition between a damaged solid and a granular phase. This breakage model, however, did not explicitly account for the underlying fundamental physics and the microscopic processes at the grain scale. Neither did the model entail spatial heterogeneities that contribute to shear localization.

\section{First and second laws of thermodynamics}\label{sec:2}

After briefly surveying experiments, simulations and theories of grain breakage in the literature, we turn now to our main theoretical development. Consider a noncrystalline system of hard grains in contact with a thermal reservoir at fixed temperature $\theta$. Let $U_T$ denote the thermal energy of this system. If the grains interact only via contact forces, they have no configurational potential energy and, therefore, no such energy is included in $U_T$. Denote by $U_G$ the surface energy of all grains; the total energy of the system is then given by $U_T + U_G$. 

Suppose that this system is driven in simple (not pure) shear by a shear stress $s$ and a pressure $p$. The first law of thermodynamics for this system is
\begin{eqnarray}
\label{eq:first_law}
 \nonumber \dot{U}_T + \dot{U}_G &=&  V s\, \dot{\gamma}^{\text{pl}} - p\, \dot{V} \\ &=&  V s\, \dot{\gamma}^{\text{pl}} - p X \dot{S}_C - p \sum_{\alpha} \left( \dfrac{\partial V}{\partial \Lambda_{\alpha}} \right)_{S_C} \dot{\Lambda}_{\alpha}, ~~~~~
\end{eqnarray}
where $\dot{\gamma}^{\text{pl}}$ is the plastic shear rate, $S_C$ is the granular configurational entropy introduced above in Eq.~(\ref{eq:compactivity_def}), and the $\Lambda_{\alpha}$ are internal variables that specify the configurational state of the granular subsystem.  

Let $S_T$ denote the entropy of the reservoir plus the (quantitatively negligible) entropy of the kinetic degrees of freedom of the grains.  Then
\begin{equation}
 \dot{U}_T = \theta \dot{S}_T ,
\end{equation}
and 
\begin{equation}\label{eq:S_C}
 p X \dot{S}_C = V s \dot{\gamma}^{\text{pl}} - p \sum_{\alpha} \left( \dfrac{\partial V}{\partial \Lambda_{\alpha}} \right)_{S_C} \dot{\Lambda}_{\alpha} - \theta \dot{S}_T - \dot{U}_G.
\end{equation}
The second law of thermodynamics requires that the total entropy be a non-decreasing function of time:
\begin{equation}\label{eq:second_law}
 \dot{S} = \dot{S}_C + \dot{S}_T \geq 0 .
\end{equation}
Substituting Eq.~\eqref{eq:S_C} for $\dot S_C$ into the second law above, and using the fact that each individually variable term in the resulting inequality must be non-negative \cite{coleman_1963,langer_2009a,langer_2009b,langer_2009c}, we arrive at the second-law constraints
\begin{eqnarray}
 \label{eq:W} {\cal W} = V s\, \dot{\gamma}^{\text{pl}} - p \sum_{\alpha} \left( \dfrac{\partial V}{\partial \Lambda_{\alpha}} \right)_{S_C} \dot{\Lambda}_{\alpha} - \dot{U}_G \geq 0; \\
 \label{eq:S_T} (p X - \theta) \dot{S}_T \geq 0 .
\end{eqnarray}
In arriving at these two constraints, we have arranged terms in such a way that terms pertaining to the degrees of freedom that belong to the same subsystem are grouped together. The dissipation rate ${\cal W}$, as defined in \cite{langer_2009a,langer_2009b,langer_2009c}, is the difference between the rate at which inelastic work is done on the configurational subsystem and the rate at which energy is stored in the internal degrees of freedom. The second constraint implies that $p X - \theta$ and $\dot{S}_T$ must carry the same sign if they are nonzero, so that
\begin{equation}\label{eq:Q}
 \theta \dot{S}_T = - {\cal K} \left( \theta - p\, X \right)\equiv \,{\cal Q},
\end{equation}
where ${\cal K}$ is a non-negative thermal transport coefficient. It is already clear from this analysis that $p\,X$ plays the role of a temperature. $p\,X$ approaches $\theta$ in an equilibrating system; and a heat flux ${\cal Q}$ flows from the granular subsystem into the reservoir when the two subsystems are not in thermodynamic equilibrium with each other.

\section{STZ equations of motion}\label{sec:3}

In this section, we repeat the derivation of the STZ equations of motion as in~\cite{lieou_2012}, but with the additional ingredient that grains may break apart, so that both the characteristic grain size $a$ and the number of particles $N$ change with time.

At this point, we make the cautionary remark that real granular systems consists of grains with many different sizes and shapes; one might thus call into question the validity of describing the comminution process in terms of a single characteristic grain size $a$ instead of tracking the evolution of the size distribution $P(a)$, which often has the form of a heavy-tail power law~\cite{sammis_1987}. An elementary calculation shows, however, that if the functional form of $P(a)$ does not change in the course of the comminution process -- as suggested by self-similar nature of the distribution, then the total surface energy $U_G$ scales with the inverse of some characteristic grain size $a$. In the following we shall therefore only refer to this characteristic grain size for simplicity.

As usual~\cite{falk_1998,langer_2011}, we suppose for simplicity that STZ's can be classified as ``plus'' and ``minus'' according to their orientations relative to the applied shear stress.  We let $N_+$ and $N_-$ denote the number of STZs in each of the two orientations, and let
\begin{equation}
 \Lambda = \dfrac{N_+ + N_-}{N(a)}; \quad m = \dfrac{N_+ - N_-}{N_+ + N_-}
\end{equation}
denote the density and orientational bias of STZ's, where $N(a)$ is the number of grains, now a function of the characteristic grain size $a$: $N(a) \propto a^{-3}$.

Let $v_Z$ denote the excess volume per STZ. Then the total volume $V$ is the sum of the constant volume $V_0 = N a^3$ of grain material plus the configurational volume associated with structural defects:
\begin{eqnarray}\label{eq:V}
 \nonumber V &=& V_0 + N \Lambda v_Z + V_1 (S_1) \\ &=& V_0 + N \Lambda v_Z + V_1 (S_C - S_Z (\Lambda, m, a) ),
\end{eqnarray}
where $V_1$ and $S_1$ are the volume and entropy of all configurational degrees of freedom of the granular system not associated with STZ's, and $S_Z$ is the entropy associated with the STZ's. Then, under the assumption the STZ's are two-state entities, we can compute $S_Z$ easily by counting the number of possible configurations of distributing $N_+$ and $N_-$ STZ's of each orientation among $N$ sites~\cite{langer_2009c}. The result is
\begin{equation}\label{eq:S_Z}
 S_Z (\Lambda, m, a) = N(a) S_0 (\Lambda) + N(a) \Lambda \psi (m)
\end{equation}
where
\begin{eqnarray}
 \label{eq:S_0} S_0 (\Lambda) &=& - \Lambda \ln \Lambda + \Lambda;\\
 \label{eq:psi}\nonumber \psi(m) &=& \ln 2 - \dfrac{1}{2} (1 + m) \ln (1 + m) \\ & & - \dfrac{1}{2} (1 - m) \ln (1 - m).
\end{eqnarray}

The STZ equation of motion for $N_+$ and $N_-$ is given by:
\begin{equation}\label{eq:master}
 \tau \dot{N}_{\pm} = {\cal R} (\pm s) N_{\mp} - {\cal R} (\mp s) N_{\pm} + \tilde{\Gamma} \left( \dfrac{1}{2} N^{\text{eq}} - N_{\pm} \right) .
\end{equation}
The corresponding strain rate is
\begin{equation}\label{eq:strainrate}
 \dot{\gamma}^{\text{pl}} = \dfrac{2\,v_0 (a)}{\tau V} \left[ {\cal R}(s) N_- - {\cal R}(-s) N_+ \right],
\end{equation}
where we define the volume of the plastic core of an STZ to be $v_0 (a)$, which ought to be proportional to $a^3$. Because we are describing simple rather than pure shear, there is a factor of $2$ up front.

In Eq.~\eqref{eq:master}, $\tau = a \sqrt{\rho_G / p}$, where $\rho_G$ denotes the material density of the grains, is the inertial time scale that characterizes the typical duration of a pressure-driven particle rearrangement event~\cite{lieou_2012,liu_2011,haxton_2012}; it is proportional to the average time between successive grain-grain collisions. This time scale also applies in a dense granular medium where inter-particle friction is important, as long as the friction is proportional to the normal force at the contact interface. Its product with the shear rate $\dot{\gamma}^{\text{pl}}$ gives the so-called inertial number, the magnitude of which determines the flow regime of dense granular flow~\cite{jop_2006}. ${\cal R}(\pm s)$ represent the rates (in units of $\tau^{-1}$) at which the STZ's are making forward and backward transitions. The term proportional to $\tilde{\Gamma}$ represents the rates of STZ creation and annihilation; $N^{\text{eq}}$ is the steady-state, total number of STZ's.  

$\tilde\Gamma/\tau$ is an attempt frequency consisting of additive thermal and mechanical parts: 
\begin{equation}
\tilde{\Gamma} = \rho +\Gamma.
\end{equation}
The quantity $\rho$ is best understood as a dimensionless, thermal noise strength. In systems composed of aggregate grains that we presently consider, $\rho$ is associated with the acoustic-vibrational motion of the grains~\cite{brodsky_2012}; vibration provides a means to un-jam a granular system so that it can explore packing configurations. In granular experiments, acoustic vibrations have been found to trigger compaction and stick-slip events~\cite{johnson_2008}, or cause a transition from a disordered to a crystalline state~\cite{daniels_2005}. When $\rho=0$, the system is fully jammed in the sense that configurational rearrangements can occur only in response to sufficiently large driving forces. If the grains are extremely fine, with size of the order of several nanometers, thermal fluctuations alone may be able to unjam the system; in such a case $\rho \neq 0$.

In analogy to $\rho/\tau$, the quantity $\Gamma/\tau$ is the contribution to the attempt frequency in Eq.~(\ref{eq:master}) due to externally applied shear. It will be computed below in Eq.~\eqref{eq:Gamma} in terms of the rate of entropy generation. 

Bearing in mind that the grain size $a$ and hence the total number of grains $N(a) = V_0 / a^3$ are now dynamical variables (the equation of motion for $a$ will be specified in the next section), the STZ equations of motion for the intensive state variables $\Lambda$ and $m$ can be written as:
\begin{eqnarray}
 \label{eq:Lambda} \tau\, \dot{\Lambda} &=& \tilde{\Gamma} ( \Lambda^{\text{eq}} - \Lambda ) + 3 \tau \dfrac{\dot{a}}{a} \Lambda; \\
 \label{eq:m} \tau\, \dot{m} &=& 2\, {\cal C}(s) ( {\cal T}(s) - m ) - \tilde{\Gamma} m - \tau \dfrac{\dot{\Lambda}}{\Lambda} m + 3 \tau \dfrac{\dot{a}}{a} m; ~~~~~ \\
 \label{eq:D_pl} \tau \,\dot{\gamma}^{\text{pl}} &=&2\, \epsilon_0\,\Lambda\, {\cal C}(s) ({\cal T}(s) - m),
\end{eqnarray}
where $\epsilon_0 = N(a)\, v_0 (a) / V_0$ is independent of $a$, and $\Lambda^{\text{eq}} = N^{eq} / N(a)$. In writing Eq.~\eqref{eq:D_pl} we have implicitly made the approximation $\Lambda \ll 1$ and $V_1 \ll V_0$ in Eq.~\eqref{eq:V} so that $V \approx V_0$ in Eq.~\eqref{eq:strainrate}. We also define
\begin{equation}
\label{eq:calC}
 {\cal C}(s) = \dfrac{1}{2} \left( {\cal R}(s) + {\cal R}(-s) \right) ;
\end{equation}
and
\begin{equation}
 {\cal T}(s) = \dfrac{{\cal R}(s) - {\cal R}(-s)}{{\cal R}(s) + {\cal R}(-s)} .
\end{equation}

\section{Grain breakage; temporal evolution of characteristic grain size}

At this point we introduce the evolution equation for the characteristic grain size $a$; it will be derived based upon basic principles and dimensional analysis alone. First, it is natural to expect that the rate at which grain comminution occurs is directly proportional to the plastic work per unit volume $\dot{\gamma}^{\text{pl}} s$ done by the imposed shear, that being the only external source of energy. This hypothesis is supported by the observation~\cite{mandl_1977} that shearing significantly reduces the compressive load necessary to fracture grains, because shear motion continuously reconfigures force chains and generates exceptionally high local contact forces which ultimately result in grain crushing. Then the evolution equation for $a$ must be of the form
\begin{equation}
 \dot{a} = - \kappa \dfrac{\dot{\gamma}^{\text{pl}} s}{u} a .
\end{equation}
Here, $\kappa$ is a dimensionless quantity that specifies the fraction of plastic work expended in the creation of new grain surfaces, and $u$ is a quantity with the dimensions of energy per unit volume. The factor $a$ on the right-hand side has been inserted for obvious dimensional reasons.

It is clear that $\kappa$ is a function of the pressure $p$. It ought to reflect the existence of a threshold confining pressure $p_{\text{th}}$ below which grain fragmentation is rare; experiments such as \cite{mair_2002} suggest that this pressure is of the order of 25 MPa. It must also reflect the fact that grain breakage below some threshold grain size rarely occurs. Thus we expect that
\begin{equation}
 \kappa = \kappa_0 \exp \left( - \dfrac{p_{\text{th}}}{p} \right),
\end{equation}
where $\kappa_0$ is a constant. The threshold pressure can be estimated using fracture mechanics as the stress necessary to fracture a grain with a characteristic flaw size proportional to its size $a$~\cite{bazant_1999}; thus
\begin{equation}
 p_{\text{th}} = \sqrt{\dfrac{2 E \gamma_G}{\pi a}},
\end{equation}
where $E$ is the Young modulus of the grain material. For quartz sand grains with typical Young modulus $E = 70$ GPa, $\gamma_G = 1$ J m$^{-2}$, and $a = 10^{-4}$ m, $p_{\text{th}} \approx 20$ MPa, roughly equal to the pressure above which grain breakage is pervasive. Notice that $\kappa \rightarrow 0$ as the pressure decreases towards zero, as it should. The characteristic grain size below which grain breakage becomes unlikely is then $a_{\text{min}} \sim 2 E \gamma_G / (\pi p^2)$.

We propose that $u$ may simply be proportional to the surface energy $\gamma_G$ per unit new area. Dimensional analysis now implies that $u = \gamma_G / a$, so that
\begin{equation}\label{eq:adot1}
 \dot{a} = - \kappa_0 \exp \left( - \dfrac{p_{\text{th}}}{p} \right) \dfrac{\dot{\gamma}^{\text{pl}} s}{\gamma_G} a^2 .
\end{equation}
Note that Eq.~\eqref{eq:adot1} involves a direct comparison of the plastic work done by the shear stress per unit volume, and the grain surface energy per unit volume. Eq.~\eqref{eq:adot1}, as written, suggests that $\dot{a}$ never vanishes as long as the plastic strain rate is nonzero; however, the dynamics of $a$ becomes exponentially slow once $a < a_{\text{min}}$, suggesting that abrasion dominates in that regime, in conformity with the simulations of Mair and Abe~\cite{mair_2011}, as we shall see below.

\section{Dissipation, heat production and mechanical noise}

\subsection{Dissipation rate and thermodynamic constraints}

At this point in the development, the second law of thermodynamics provides useful constraints on the ingredients of the equations of motion. If $\gamma_G$ denotes the surface energy per unit area of the grains, then the total surface energy $U_G$, which we first introduced in Eq.~\eqref{eq:first_law}, equals
\begin{equation}\label{eq:U_G}
 U_G = \gamma_G N(a) a^2 = \gamma_G V_0 / a .
\end{equation}
We now substitute this and Eqs.~\eqref{eq:Lambda}, \eqref{eq:m}, \eqref{eq:D_pl} and~\eqref{eq:adot1} into Eq.~\eqref{eq:W} for the dissipation rate, keeping in mind that the extensive volume $V$ now depends on $a$ through the $N$-dependence in $S_1 = S_C - S_Z (\Lambda, m, a)$, and that the excess volume $v_Z$ of each STZ ought to be proportional to the typical grain volume: $v_Z = \epsilon_Z a^3$ for some constant $\epsilon_Z$. We also use the approximation $V \approx V_0$ where appropriate. The result is
\begin{eqnarray}\label{eq:W2}
 \nonumber \tau\, \dfrac{{\cal W}}{N(a)} &=& - \tilde{\Gamma}\, p\, X\, \Lambda\, m \dfrac{d \psi}{dm} - p\, \tilde{\Gamma}\,(\Lambda^{\text{eq}} - \Lambda) \\\nonumber && \times \left[ \epsilon_Z a^3 + X \left( \ln \Lambda - \psi (m) + m \,\dfrac{d \psi}{dm} \right) \right] \\ && + 2\, \Lambda\, {\cal C}(s)\,\Bigl( {\cal T}(s) - m\Bigr) \left( {\cal A} \epsilon_0 a^3 s + p X \dfrac{d \psi}{dm} \right)~~~~~,
\end{eqnarray}
where
\begin{equation}
 {\cal A} = 1 - \kappa_0 \exp \left(- \dfrac{p_{\text{th}}}{p} \right) \left[ 1 - \dfrac{3 p a}{\gamma_G} \epsilon_Z \left( \dfrac{X}{\epsilon_Z a^3} + 1 \right) \Lambda \right] .
\end{equation}
The second-law constraint, ${\cal W} \geq 0$, must be satisfied by all possible motions of the system; this is guaranteed if each of the three terms in Eq.~\eqref{eq:W2} is non-negative~\cite{coleman_1963,langer_2009a,langer_2009b,langer_2009c}. (Indeed, only the third term depends explicitly on the shear stress $s$, while the second term is proportional to $\dot{\Lambda}$; the entire expression must be non-negative irrespective of $s$ and $\dot{\Lambda}$.) The first term automatically satisfies this requirement because, from Eq.~\eqref{eq:psi}, we have
\begin{equation}\label{eq:dpsidm}
 \dfrac{d \psi}{dm} = - \dfrac{1}{2} \ln \left( \dfrac{1 + m}{1 - m} \right) = - \tanh^{-1} (m)
\end{equation}
so that the product $- m ( d \psi / dm)$ is automatically non-negative. 

The non-negativity constraint on the second term in Eq.~\eqref{eq:W2} can be written in the form
\begin{equation}
 - \dfrac{\partial F}{\partial \Lambda} (\Lambda^{\text{eq}} - \Lambda) \geq 0
\end{equation}
where $F$ is a free energy given by
\begin{equation}
 F (\Lambda, m) = p \left[ \epsilon_Z a^3 \Lambda - X S_0 (\Lambda) - X \Lambda \left( \psi(m) - m \dfrac{d \psi}{d m} \right) \right].
\end{equation}
$\Lambda^{\text{eq}}$ must be the value of $\Lambda$ at which $ \partial F / \partial \Lambda$ changes sign, so that
\begin{equation}\label{eq:Lambda_eq}
 \Lambda^{\text{eq}} = \exp \left[ - \dfrac{\epsilon_Z a^3}{X} + \psi(m) - m \dfrac{d \psi}{dm} \right] \approx 2\, \exp \left( - \dfrac{\epsilon_Z a^3}{X} \right). ~~~~~
\end{equation}
Thus, the STZ density in this non-equilibrium situation is given by a Boltzmann-like expression in which the compactivity plays the role of the temperature.

As for the third term, we have
\begin{equation}\label{eq:Ts}
 \Bigl( {\cal T}(s) - m \Bigr) \left( {\cal A} \epsilon_0 a^3 s + p X \dfrac{d \psi}{dm} \right) \geq 0.
\end{equation}
The two factors on the left-hand side must be monotonically increasing functions of $s$ that change sign at the same point for arbitrary values of $m$. According to Eq.~\eqref{eq:dpsidm}, this is possible only if
\begin{equation}
 {\cal T}(s) = \tanh \left( \dfrac{{\cal A} \epsilon_0 a^3 s}{p X} \right).
\end{equation}

\subsection{Heat production and quasistationary relations}

Observe that the equations of motion for $\Lambda$ and $m$, Eqs.~\eqref{eq:Lambda} and~\eqref{eq:m}, contain no factor of $\Lambda \ll 1$. On the other hand, $\dot{a}$ is directly proportional to the plastic strain rate $\dot{\gamma}^{\text{pl}}$ which is proportional to $\Lambda$. Thus $\Lambda$ and $m$ are fast variables while $a$ is a slow variable; we therefore set $\dot{\Lambda} = \dot{m} = 0$, implying specifically that $\Lambda = \Lambda^{\text{eq}}$. We start by assuming that Pechenik's hypothesis \cite{langer_2009c,pechenik_2003} remains valid for a sheared granular material; that is, that the mechanical noise strength $\Gamma$ is proportional to the mechanical work per STZ. The plastic work per unit volume is simply $\dot{\gamma}^{\text{pl}} s$. To convert this rate into a noise strength with dimensions of inverse time, we multiply by the volume per STZ, $V_0 / (N \Lambda^{\text{eq}})$, and divide by an energy conveniently written in the form $\epsilon_0 (V_0 / N)\, s_0$. Here, $s_0$ is a system-specific parameter with the dimensions of stress. The resulting expression for $\Gamma$ is
\begin{equation}\label{eq:Gamma}
 \Gamma = \dfrac{\tau \dot{\gamma}^{\text{pl}} s}{\epsilon_0 s_0 \Lambda^{\text{eq}}} = \dfrac{2 s}{s_0} {\cal C}(s) \Bigl( {\cal T}(s) - m \Bigr).
\end{equation}
With this result, the stationary version of Eq.~\eqref{eq:m} reads
\begin{equation}
 2\, {\cal C}(s) \Bigl( {\cal T}(s) - m \Bigr) \left( 1 - \dfrac{m s}{s_0} \right) - m\, \rho  = 0 .
\end{equation}
The stationary value of $m$ is then given by
\begin{eqnarray}\label{eq:m_eq}
 \nonumber m^{\text{eq}} (s) &=& \dfrac{s_0}{2s} \left[ 1 + \dfrac{s}{s_0} {\cal T}(s) + \dfrac{\rho}{2 {\cal C}(s)} \right] \\ &&- \dfrac{s_0}{2s} \sqrt{\left[ 1 + \dfrac{s}{s_0} {\cal T}(s) + \dfrac{\rho}{2 {\cal C}(s)} \right]^2 - 4 \dfrac{s}{s_0} {\cal T}(s)}. ~~~~~
\end{eqnarray}
In particular, when $\rho = 0$, we find
\begin{equation}
 m^{\text{eq}} = 
\begin{cases}
{\cal T}(s), & \text{if $(s / s_0)\, {\cal T}(s) < 1$} , \\
s_0/s, & \text{if $(s / s_0)\, {\cal T}(s) \geq 1$}.
\end{cases}
\end{equation}
Thus an exchange of stability occurs in a similar manner as it did in systems where constituent particles do not break apart; the yield stress for a completely jammed system is the solution of the equation
\begin{equation}
 s_y \, {\cal T}(s_y) = s_y \tanh \left( \dfrac{{\cal A} \epsilon_0 a^3 s_y}{p X}\right) = s_0.
\end{equation}
If the temperature-like quantity $p\,X$ is small in comparison with ${\cal A} \epsilon_0 a^3 s_0$, then $s_y \approx s_0$. For practical purposes, ${\cal A} \epsilon_0 a^3 s_0 / (p X) \ll 1$ ought to hold whenever $(s / s_0 ) {\cal T} (s) \geq 1$; thus we may as well take ${\cal T} (s) \approx 1$. Then $s_0$ sets, in effect, the minimum flow stress of the system.

Finally, the steady-state version of Eq.~\eqref{eq:D_pl} for the strain rate becomes
\begin{equation}\label{eq:q_s}
 q \equiv \tau \dot{\gamma}^{\text{pl}} = 4\,\epsilon_0\,e^{-\,1/\chi}\,{\cal C}(s) \left[1 - m^{\text{eq}} (s) \right] .
\end{equation}
Here, we have introduced the inertial number $q$ as a dimensionless measure of the strain rate. (See for example \cite{jop_2006}.)   We also have introduced the dimensionless compactivity $\chi = X/v_Z$.

\section{Compactivity: kinematics and thermodyanmics}

The dimensionless compactivity $\chi$, introduced above, measures the amount of configurational (i.e.~ structural) disorder in the granular system. When the system is completely jammed, i.e.~$\rho = 0$, externally applied shear constitute the only means to ``stir'' the system and cause granular rearrangement; in such a case, the steady-state compactivity ought to be a function of the strain rate alone: $\chi = \hat{\chi} (q)$. It should approach some constant in the limit of small $q$, and should become a rapidly increasing function of $q$ once the shear rate becomes comparable to the rate of intrinsic structural relaxation. A deeper discussion of the quantity $\hat{\chi} (q)$ can be found in~\cite{lieou_2012,haxton_2007,manning_2007b}; for our purposes it suffices to write the inverse relation $q (\hat{\chi})$ in the Vogel-Fulcher-Tamann (VFT) form
\begin{equation}\label{eq:vft}
 \dfrac{1}{q} = \dfrac{1}{q_0} \exp \left[ \dfrac{A}{\hat{\chi}} + \alpha_{\text{eff}} (\hat{\chi}) \right] ,
\end{equation}
where
\begin{equation}
 \alpha_{\text{eff}} (\hat{\chi}) = \left( \dfrac{\hat{\chi}_1}{\hat{\chi} - \hat{\chi}_0} \right) \exp \left( - b \dfrac{\hat{\chi} - \hat{\chi}_0}{\hat{\chi}_A - \hat{\chi}_0} \right) .
\end{equation}

The quantity $\chi$ evolves according to the first law of thermodynamics. Remembering the $\chi = X / (\epsilon_Z a^3)$ is dimensionless and dependent on the time-dependent grain size $a$, and that the compactivity $X = \partial V / \partial S_C$, having the dimensions of volume, is the more fundamental temperature-like quantity, we see that it is necessary to return to the extensive variable for the time being, to deduce the equation of motion for the compactivity based on concepts of energy flow. To this end, note that
\begin{eqnarray}\label{eq:Xdot1}
 \nonumber X \dot{S}_C &=& X \sum_{\alpha} \left( \dfrac{\partial S_C}{\partial \Lambda_{\alpha}} \right)_X \dot{\Lambda}_{\alpha} + X \left( \dfrac{\partial S_C}{\partial X} \right)_{ \{ \Lambda_{\alpha} \} } \dot{X} \\ & \equiv & C^{\text{eff}} \dot{X} + X \sum_{\alpha} \left( \dfrac{\partial S_C}{\partial \Lambda_{\alpha}} \right)_X \dot{\Lambda}_{\alpha}.
\end{eqnarray}
Here $C^{\text{eff}} \equiv X (\partial S_C / \partial X)_{ \{ \Lambda_{\alpha} \} } = (\partial V / \partial X)_{ \{ \Lambda_{\alpha} \} }$ can be interpreted as an effective volume expansion coefficient. Because $V$ is independent of the number of grains $N \sim a^{-3}$, and $S_C \sim N \sim a^{-3}$, we have $X \sim a^3$, as it should (see also the remark following Eq.~\eqref{eq:q_s}), so that $C^{\text{eff}} \sim a^{-3}$. Then
\begin{eqnarray}\label{eq:Xdot2}
 \nonumber C^{\text{eff}} \dot{X} &=& X \dot{S}_C - X \sum_{\alpha} \left( \dfrac{\partial S_C}{\partial \Lambda_{\alpha}} \right)_X \dot{\Lambda}_{\alpha} \\ \nonumber &=& V \dfrac{s}{p} \dot{\gamma}^{\text{pl}} - \sum_{\alpha} \left( \dfrac{\partial V}{\partial \Lambda_{\alpha}} \right)_{S_C} \dot{\Lambda}_{\alpha} - \dfrac{\theta}{p} \dot{S}_T \\ \nonumber & & + \dfrac{V_0 \gamma_G}{p a^2} \dot{a} - X \sum_{\alpha} \left( \dfrac{\partial S_C}{\partial \Lambda_{\alpha}} \right)_X \dot{\Lambda}_{\alpha} \\ &=& V \dfrac{s}{p} \dot{\gamma}^{\text{pl}} - {\cal K} \left( X - \dfrac{\theta}{p} \right) + V_0 \dfrac{\gamma_G}{p a^2} \dot{a}.
\end{eqnarray}
In arriving at the last expression, we used Eq.~(4.2), $V = V_0 + \epsilon_Z \Lambda + V_1 (S_C - S_Z (\Lambda, m, a))$, to compute $(\partial V / \partial m)_{S_C} = - X \partial S_Z / \partial m$, $(\partial V / \partial a)_{S_C} = - X \partial S_Z / \partial a$, and $(\partial V / \partial \Lambda)_{S_C} = \epsilon_Z V_0 - X \partial S_Z / \partial \Lambda$. We also used the fact that $S_C = S_Z (\Lambda, m, a) + S_1$, with $S_1$ being indepedent of the $\Lambda_{\alpha}$ degrees of freedom, to write $(\partial S_C / \partial \Lambda_{\alpha} )_X = \partial S_Z / \partial \Lambda_{\alpha}$. Furthermore, we used the smallness of $\Lambda$ to drop the residual term proportional to $\dot{\Lambda}$. ${\cal K}$ is the thermal transport coefficient that first appeared in Eq.~\eqref{eq:Q}. 

We now specialize to athermal situations for which we may set $\theta = 0$, and compute ${\cal K}$ based on knowledge about the steady state in $\chi$: $\chi = \hat{\chi} (q)$. Because $X = \epsilon_Z a^3 \chi$, $\dot{\chi} = 0$ does not imply $\dot{X} = 0$; rather,
\begin{equation}
 \dot{X} = \epsilon_Z a^3 \left( \dot{\chi} + \dfrac{3 \dot{a}}{a} \chi \right).
\end{equation}
With this in mind, we substitute $X = \hat{X} = \epsilon_Z a^3 \hat{\chi} (q)$ into Eq.~\eqref{eq:Xdot2} and set $\dot{\chi} = 0$ there. It follows that
\begin{equation}
 {\cal K} = \dfrac{V (s/p) \dot{\gamma}^{\text{pl}} - C^{\text{eff}} \epsilon_Z a^3 (3 \dot{a} / a) \hat{\chi} + V_0 (\gamma_G / p a^2) \dot{a}}{\epsilon_Z a^3 \hat{\chi} (q)}.
\end{equation}
The equation of motion for the dimensionless compactivity $\chi$ is therefore
\begin{eqnarray}\label{eq:chidot}
 \nonumber \dot{\chi} &=& \left( 1 - \dfrac{\chi}{\hat{\chi} (q)} \right) \left[ \dfrac{V_0 (s/p) \dot{\gamma}^{\text{pl}}}{C^{\text{eff}} \epsilon_Z a^3} \left( 1 - \kappa_0 e^{- p_{\text{th}} / p} \right) \right] \\ &=& \left( 1 - \dfrac{\chi}{\hat{\chi} (q)} \right) \left( \dfrac{s}{p} \right) \dot{\gamma}^{\text{pl}} \left( \dfrac{1 - \kappa_0 e^{- p_{\text{th}} / p}}{c_0 \epsilon_Z} \right).
\end{eqnarray}
Here, we defined $c_0 \equiv C^{\text{eff}} a^3 / V_0$ following the remark after Eq.~\eqref{eq:Xdot1}.

\section{Theoretical predictions: stress-strain response, energy budget and grain size evolution}

Thus far we have specified the internal, microscopic dynamics through Eqs.~\eqref{eq:Lambda}, ~\eqref{eq:m}, and~\eqref{eq:adot1} for the internal variables $\Lambda$, $m$ and $a$; Eq.~\eqref{eq:chidot} for the dimensionless compactivity $\chi$; and Eq.~\eqref{eq:D_pl} for the plastic strain rate $\dot{\gamma}^{\text{pl}}$. To complete the dynamical description of the system, they are supplemented with an equation for the temporal evolution of the shear stress of the form
\begin{equation}\label{eq:sdot}
 \dot{s} = G (\dot{\gamma} - \dot{\gamma}^{\text{pl}}).
\end{equation}
Here, $\dot{\gamma}$ is the total, imposed strain rate, and $G$ is the aggregate shear modulus of the granular material~\cite{brodsky_2012}. Equation~\eqref{eq:sdot} implicitly assumes linear elasticity, approximately valid in the long-wavelength limit~\cite{makse_2004,walton_1987}; thus we adopt it here.

%%%%%%%%%%%%% FIGURE 2 %%%%%%%%%%%%
\begin{figure}[ht]
\centering 
\subfigure{\includegraphics[width=.50\textwidth]{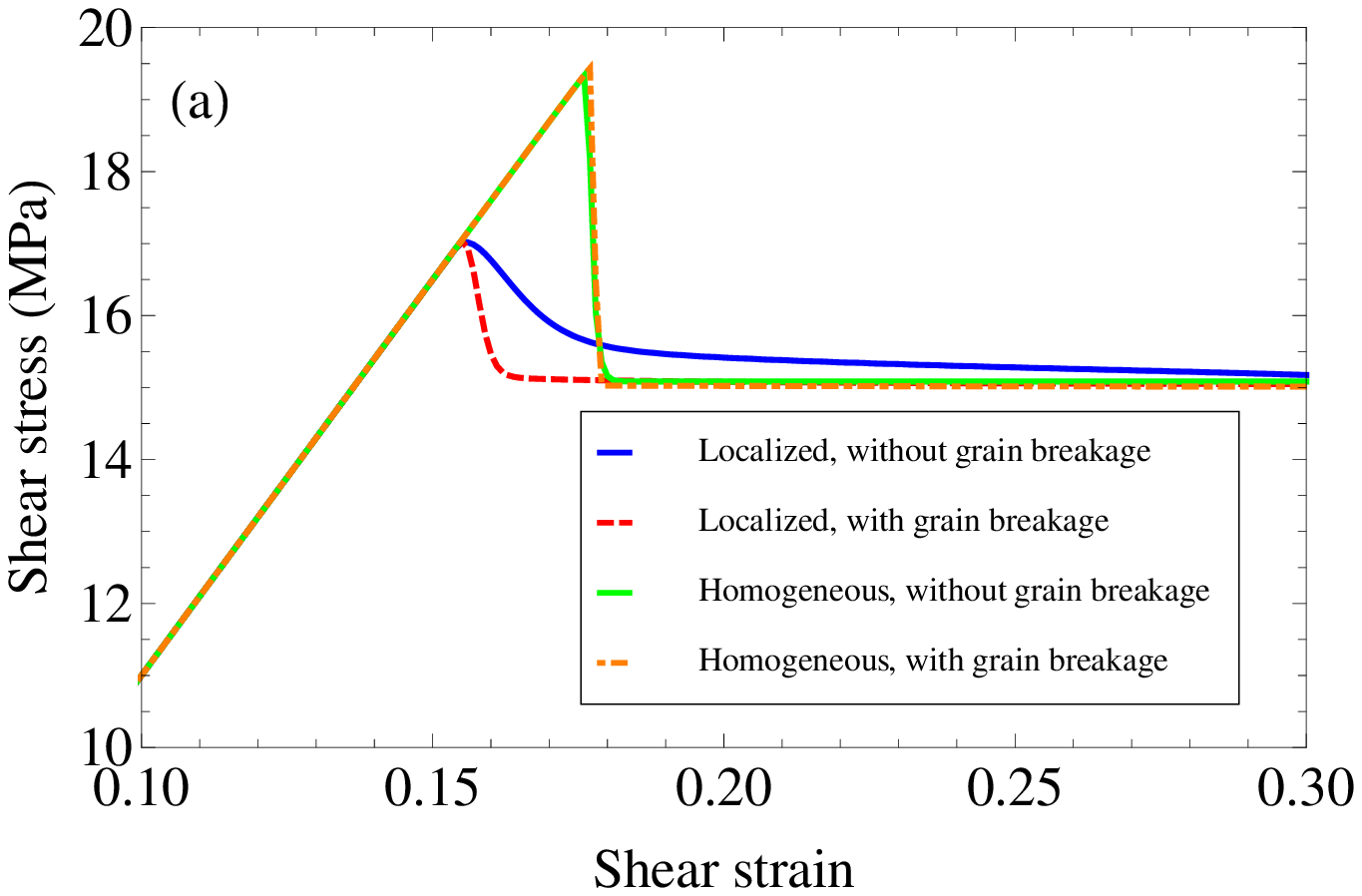}\label{fig:sxplotf51}}
\subfigure{\includegraphics[width=.50\textwidth]{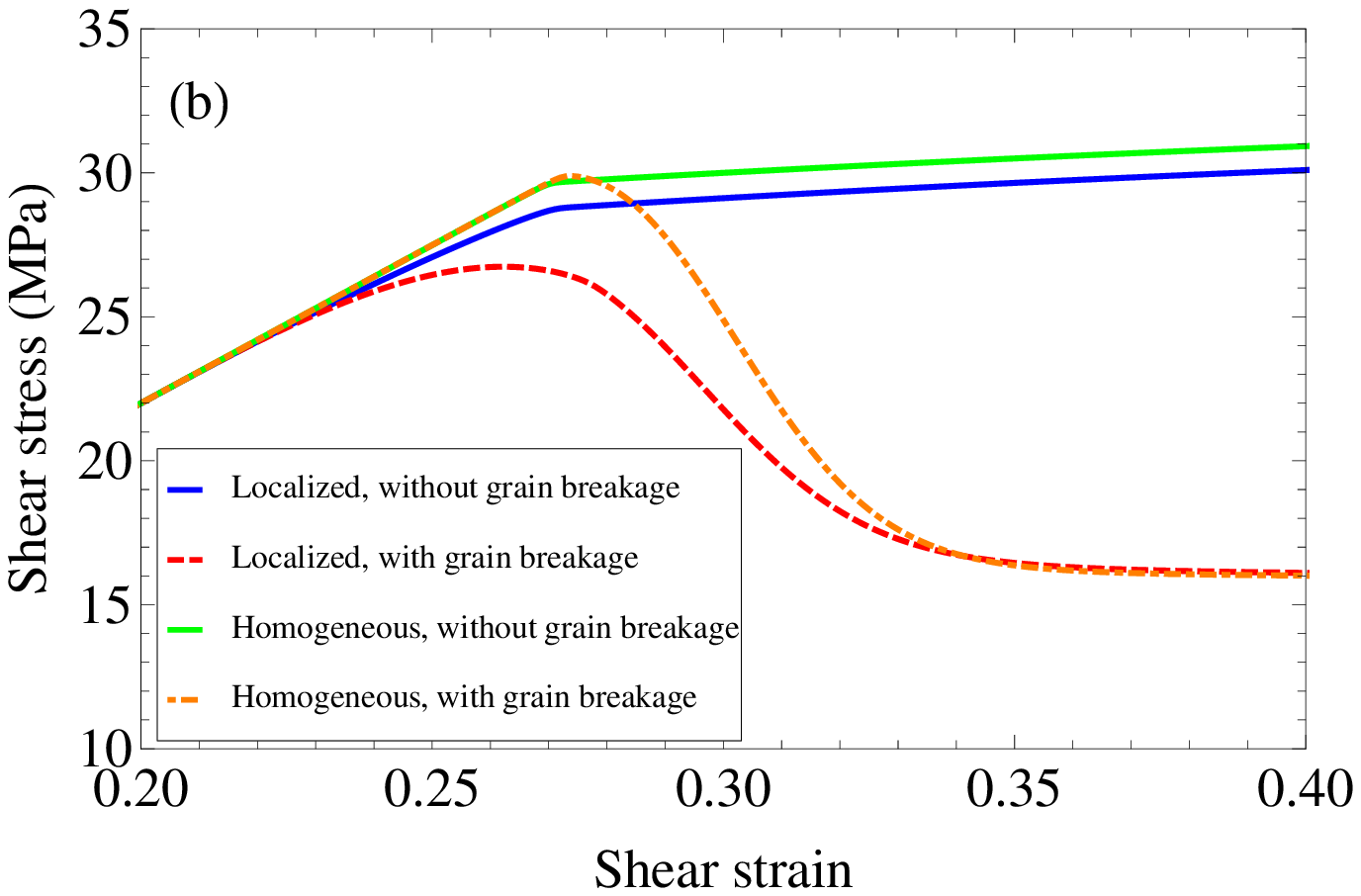}\label{fig:sxplotf65}} \caption{\label{fig:sxplot}(Color online) Variation of the shear stress $s$ with the accumulated strain, in the presence of a shear band, with and without grain fragmentation (red dotted curve and blue solid curve, respectively); and in the absence of strain localization, with and without grain fragmentation (orange dot-dashed curve and green solid curve, respectively). (a) The initial compactivity equals $\chi(t = 0) = 0.055$ and the imposed strain rate is $\dot{\gamma} = 0.1$ s$^{-1}$; a disorder-limited shear band forms when we add a small inhomogeneity to $\chi(t = 0)$ (see the next section). (b) The initial compactivity equals $\chi(t = 0) = 0.065$ and the imposed strain rate is $\dot{\gamma} = 10$ s$^{-1}$; a diffusion-limited shear band forms when we add a small inhomogeneity to $\chi(t = 0)$ (see the next section). Grain fragmentation significantly reduces the flow stress at a fast shear rate. In both cases, the softening is enhanced by strain localization. We show only the stress-strain curves in the vicinity of the onset of plastic deformation for clarity; the shear stress increases linearly with the shear strain in the elastic regime.} 
\end{figure}
%%%%%%%%%%%%%%%%%%%%%%%%%%%%%%%%%%%%%

We are now in a position to predict implication of grain breakage on flow rheology. The orange, dot-dashed curves in Figs.~\ref{fig:sxplotf51} and~\ref{fig:sxplotf65} shows the evolution of the shear stress $s$ with accumulated shear strain $\gamma = \dot{\gamma} t$ for a granular material subject to grain fragmentation, at two different shear rates $\dot{\gamma} = 0.1$ s$^{-1}$ and $10$ s$^{-1}$, respectively. The green, solid curves show the stress-strain relation when grain fragmentation does not occur; to compare the flow rheology in the presence and absence of grain comminution, we set $\kappa_0 = 0.1$ and $0$ in each of the two cases. The blue, solid curves and the red, dotted curves depict the stress-strain relation when strain localization occurs -- that is, when there is spatial heterogeneity within the material. We will return to a discussion of the implication of grain breakage on strain localization in the next section.

Here, the material density of each grain is $\rho_G = 1600$ kg m$^{-3}$, typical of quartz sand, and the typical initial grain size is $a (t = 0) = 0.1$ mm. The material is sheared at a strain rate of $\dot{\gamma} = 0.1$ s$^{-1}$, at fixed confining pressure $p = 25$ MPa; the initial conditions for the shear stress and the compactivity are $s (t = 0) = 0.01$ Pa, $\chi (t = 0) = 0.055$ at the slow shear rate of $\dot{\gamma} = 0.1$ s$^{-1}$ in Fig.~\ref{fig:sxplotf51}, and $\chi (t = 0) = 0.065$ at the fast shear rate of $\dot{\gamma} = 10$ s$^{-1}$ in Fig.~\ref{fig:sxplotf65}, when deformation is spatially homogeneous. In the case of localized deformation, $\chi (t = 0)$ has a small spike about the center of the material. The surface energy per unit area is $\gamma_G = 1$ J m$^{-2}$, the Young modulus of each grain is $E = 70$ GPa, and the aggregate shear modulus is $G = 110$ MPa, again typical of quartz sand.

The imposed strain rate in Fig.~\ref{fig:sxplotf51}, $\dot{\gamma} = 0.1$ s$^{-1}$, is slow; as such, grain breakage has no discernible effect on the stress-strain behavior when deformation is homogeneous. When strain localization is present, however, grain breakage causes the shear stress to approach the steady-state value faster. Grain breakage is therefore a softening mechanism in the presence of strain localization, which is typical in the shear flow of gouge material. In contrast, Fig.~\ref{fig:sxplotf65}, where the shear rate $\dot{\gamma} = 10$ s$^{-1}$ is two orders of magnitude faster, shows that grain fragmentation results in a significant reduction in the flow stress, irrespective of the occurrence of strain localization. We find that the flow stress is largely determined by the strain rate. On the other hand, material preparation and aging, reflected by $\chi(t = 0)$, determines the peak stress at slow shear rates; the peak stress increases as $\chi(t = 0)$ decreases, i.e., when the granular material has undergone a greater amount of aging. However, $\chi(t = 0)$ has important implications on the type of strain localization that may occur; this will be discussed in the next section.

In computing the theoretical curves in Fig.~\ref{fig:sxplot} we have made a number of simplifications to the theory above. Firstly, we assumed, as in our analysis of sheared hard spheres~\cite{lieou_2012}, that the STZ transition rate factor ${\cal C} (s)$ is insensitive to variations of the shear stress $s$; thus ${\cal C} (s) = R_0$ for some constant $R_0$. The spirit of this simplification is to permit us capture the key qualitative features of the effect of grain fragmentation, without specifying model-specific ingredients such as deformation by rolling and slipping, that the rate factor might depend on. (The only requirement on ${\cal C}(s)$ is that it is an even function of $s$. The present approximation should be valid as long as $s$ is not very much bigger than the pressure $p$.) We know from~\cite{lieou_2012} that the STZ transition rate in the jammed phase, when measured in units of the inverse time scale $\tau^{-1}$ (here our parameter choices imply that $\tau \approx 8 \times 10^{-7}$ s), is very much smaller than unity, a hallmark of the so-called ``glassy slowing-down''; thus we have somewhat arbitrarily chosen $R_0 = 0.01$. Secondly, for small shear rates $q = \tau \dot{\gamma}$, $\hat{\chi} (q)$ is not very sensitive to variations of $q$; thus $\hat{\chi} (q) \approx \hat{\chi}_0$. Setting $\chi \rightarrow \hat{\chi}_0$ in the equation for the dimensionless strain rate, Eq.~\eqref{eq:q_s}, we see that this simplification amounts to the assumption that the material is rate-strengthening. We took $\hat{\chi}_0 = 0.12$ following past experience~\cite{lieou_2012,manning_2007b}. As another simplification, we assumed that $s_0$ is independent of the grain size $a$, and chose $p = 25$ MPa and $s_0 = 15$ MPa. This has been done so that the minimum macroscopic friction at steady flow for both systems equals $s_0 / p = 0.6$, in conformity with the apparent independence of friction on grain breakage at slow shear rates, as seen in~\cite{mair_2002} for angular grains. The other parameter values used in this calculation are $c_0 = 0.01$, $\epsilon_0 = 1.5$, and $\epsilon_Z = 0.5$. The values of $\epsilon_0$ and $\epsilon_Z$ are chosen such that the plastic core volume and the excess volume associated with each STZ should be slightly bigger and smaller, respectively, than the volume of each grain~\cite{lieou_2012}, while the value of $c_0$ has been chosen such that the volume compaction is about $3.5\%$ at a shear strain of $0.5$ (see the discussion after Fig.~\ref{fig:sxdispplot} below). The amount of compaction caused by grain comminution is not reported in many experiments and simulations on the subject, but our result is in apparent conformity with the distinct element simulations of~\cite{guo_2006}, which showed roughly the same amount of volume compaction. The availability of further data could impose useful constraints on our model. Other parameter values have been chosen to illustrate the key qualitative features of our predictions in a clear manner. We summarize our choice of parameters in Table~\ref{tab:parameters}.

\begin{table}
\caption{\label{tab:parameters}List of parameters in our numerical integration of the STZ equations}
\begin{center}
%\begin{tabular*}{.45\textwidth}{ccc}
\begin{tabular}{ >{\centering\arraybackslash}m{.08\textwidth} >{\centering\arraybackslash}m{.29\textwidth} >{\centering\arraybackslash}m{.08\textwidth}}
\hline
Parameter & Description & Value \\
\hline
$s_0$ & Minimum flow stress & 25 MPa \\
$p$ & Confining pressure & 15 MPa \\
$E$ & Young modulus of each individual grain & 70 GPa \\
$G$ & Aggergate shear modulus & 110 MPa \\
$\gamma_G$ & Surface energy per unit area & 1 J m$^{-2}$ \\
$\rho_G$ & Material density of each individual grain & 1600 kg m$^{-3}$ \\
$\kappa_0$ & Intrinsic grain breakage rate & 0.1 \\
$R_0$ & Characteristic STZ transition rate & 0.01 \\
$c_0$ & Dimensionless effective volume expansion coefficient & 0.01 \\
$\hat{\chi}_0$ & Steady-state dimensionless compactivity & 0.12 \\
$\epsilon_0$ & Plastic core volume per STZ in units of grain volume & 1.5 \\
$\epsilon_Z$ & Excess volume per STZ in units of grain volume & 0.5 \\
\hline
\end{tabular}
\end{center}
\end{table}

At the granular scale, the relevant measure of strain rate $q \equiv \tau \dot{\gamma}^{\text{pl}}$ must be in units of the attempt frequency $\tau^{-1}$ which, as has been pointed out by us and in other references~\cite{lieou_2012,liu_2011,haxton_2012}, is given by $\tau = a \sqrt{ \rho_G / p}$ (see the remark following Eq.~\eqref{eq:strainrate}). Therefore, for a given $\dot{\gamma}$, grain fragmentation in fact lowers the strain rate $q$ measured in units of the attempt frequency. As we have pointed out, the simplification that $\hat{\chi} (q) \approx \hat{\chi}_0$ implies that the material is rate-strengthening; that is, the flow stress increases monotonically with the dimensionless shear rate $q$ when it is large enough. This explains why in our numerical calculations, grain breakage reduces the flow stress at a fast enough imposed shear rate $\dot{\gamma}$. On the other hand, if the material is rate-weakening -- this may be modeled by using the full Vogel-Fulcher form, Eq.~\eqref{eq:vft}, for the steady-state compactivity $\hat{\chi}$ and choosing $A < 1$~\cite{daub_2009} -- then grain fragmentation may result in an increase in the flow stress at large shear rates.

%%%%%%%%%%%%% FIGURE 3 %%%%%%%%%%%%
\begin{figure}[here]
\centering 
\subfigure{\includegraphics[width=.5\textwidth]{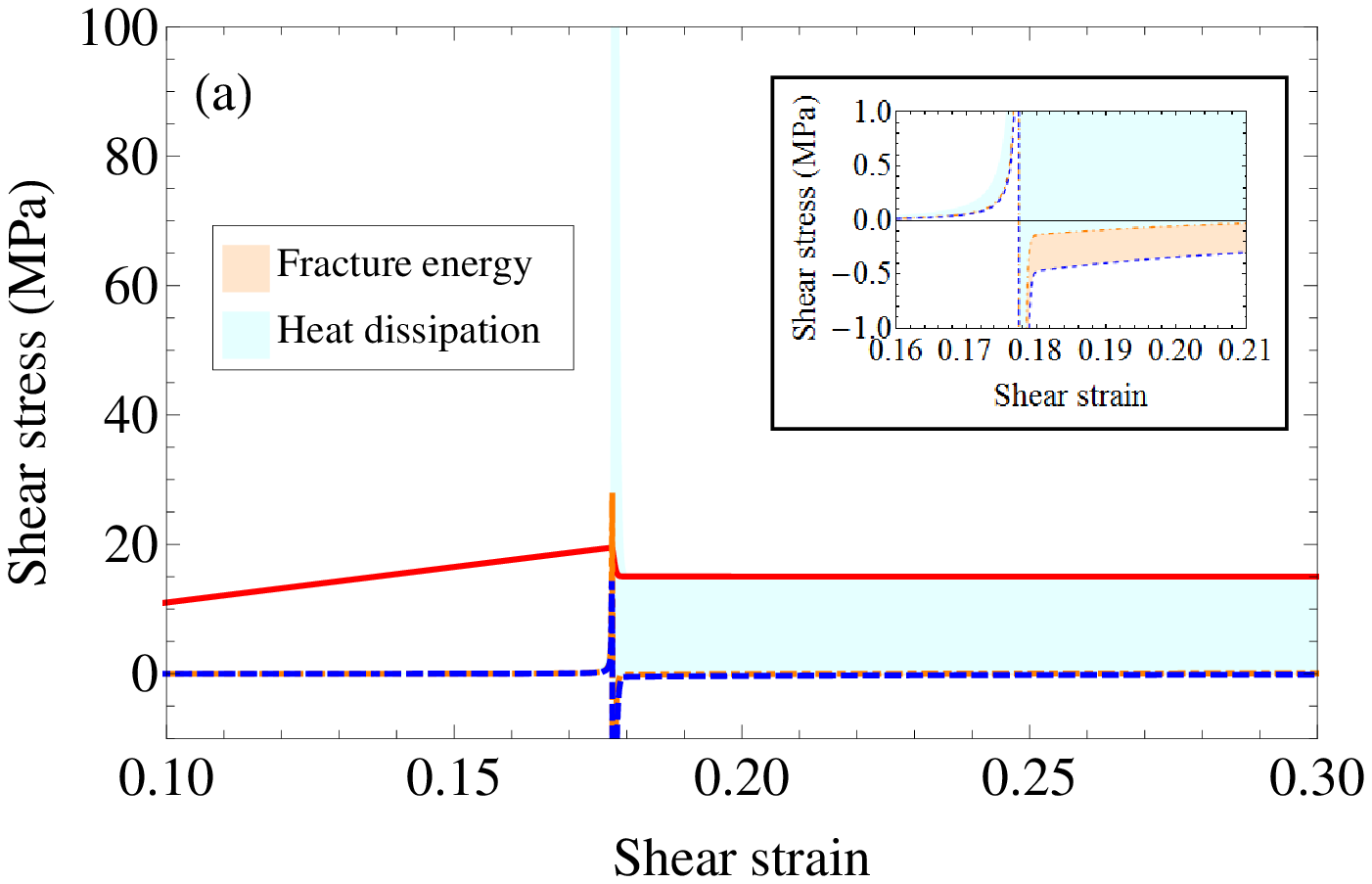}\label{fig:sxdispplotf51}}
\subfigure{\includegraphics[width=.5\textwidth]{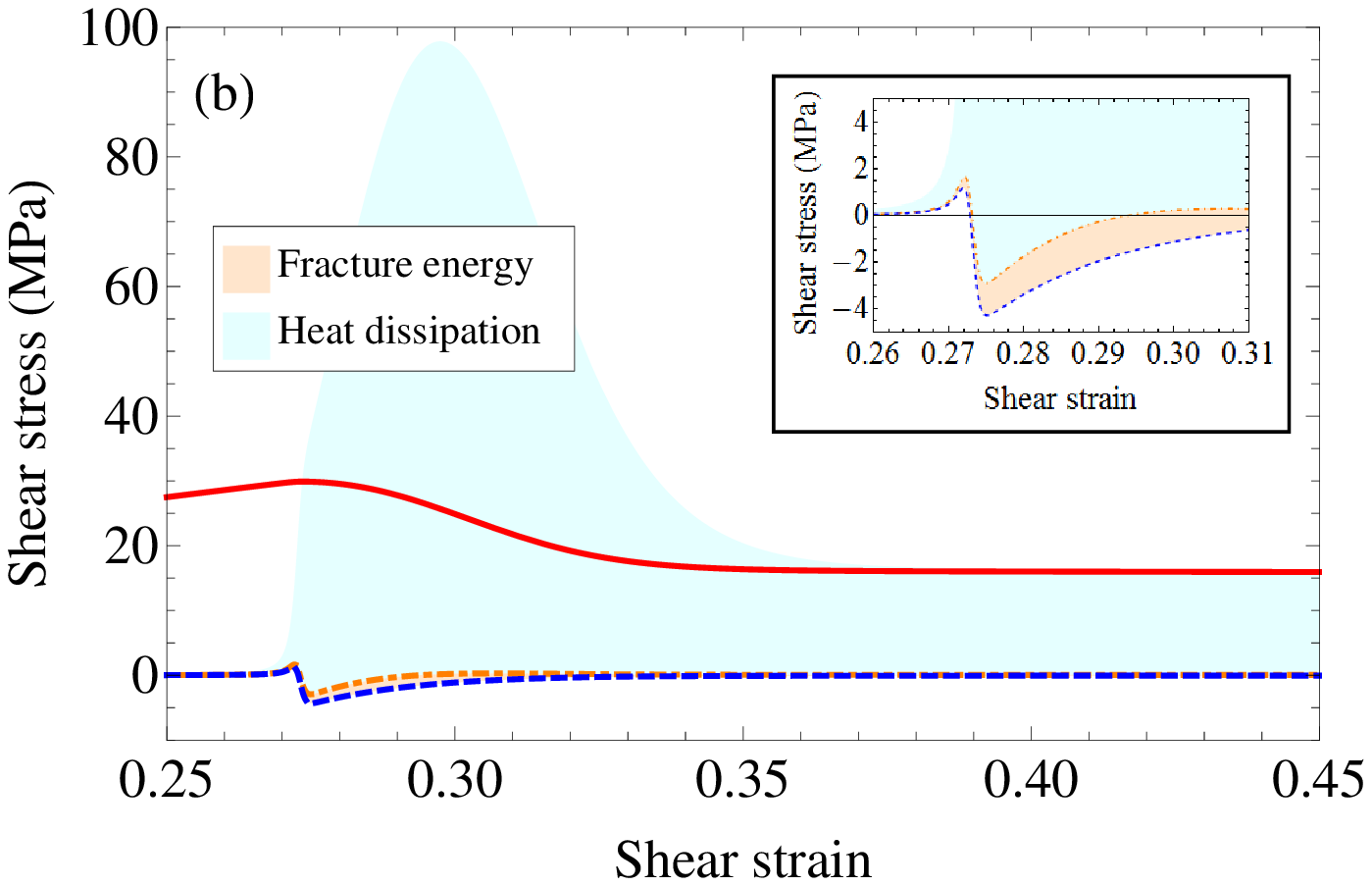}\label{fig:sxdispplotf65}}
\caption{\label{fig:sxdispplot}(Color online) Variation of shear stress with accumulated shear strain (red solid curve), work done against the confining pressure per unit volume per unit strain $p \dot{V} / (V_0 \dot{\gamma})$ (blue dashed curve), and the sum of the work done against the confining pressure and the fracture energy release rate per unit volume per unit strain $(p \dot{V} + \dot{U}_G) / (V_0 \dot{\gamma})$ (orange dot-dashed curve), when (a) $\dot{\gamma} = 0.1$ s$^{-1}$ and $\chi(t = 0) = 0.055$, and (b) $\dot{\gamma} = 10$ s$^{-1}$ and $\chi(t = 0) = 0.065$. In each of the two figures, the area between the magenta solid and orange dot-dashed curves, shaded in light orange, represents the ``fracture energy'' per unit volume, or the amount of energy per unit volume that goes into the creation of new grain surfaces following grain comminution. According to Eq.~\eqref{eq:first_law}, the area shaded in light cyan, enveloped from above by the volume-specific plastic work per unit strain $s \dot{\gamma}^{\text{pl}} / \dot{\gamma}$, represents the heat dissipation per unit volume. The energy consumed by grain breakage goes to zero at large strains. This indicates that granular rearrangements dominate over grain fragmentation at later stages when the grains become small. We show only the results in the vicinity of the onset of irreversible plastic deformation for clarity; the heat dissipation within the elastic regime equals zero.}
\end{figure}
%%%%%%%%%%%%%%%%%%%%%%%%%%%%%%%%%%%%% 

Figures~\ref{fig:sxdispplotf51} and~\ref{fig:sxdispplotf65} show the partitioning of external work into fracture energy, heat dissipation and the work of volume compaction, for the two sets of initial conditions depicted in Figs.~\ref{fig:sxplotf51} and~\ref{fig:sxplotf65} respectively. In each of the two figures here, the red solid curve shows the stress-strain response; the area shaded in light orange depicts the fracture energy per unit volume. It is straightforward to compute the ``fracture energy'', or the amount of energy absorbed by grain breakage: the rate at which energy is stored per unit volume in newly-created grain surfaces as a result of comminution is simply $\dot{U}_G / V_0 = - (\gamma_G / a^2) \dot{a} = \kappa_0 \exp ( - p_{\text{th}} / p ) \dot{\gamma}^{\text{pl}} s$. The time integral of this quantity gives the fracture energy per unit volume. Within our choice of parameters, fracture energy amounts to roughly $3 \%$ of external work at small strains in both cases. The heat dissipation can be computed from the first law of thermodynamics, Eq.~\eqref{eq:first_law}, according to which $\dot{U}_T = s \dot{\gamma}^{\text{pl}} - \dot{U}_G - p \dot{V}$. The dashed blue curve shows the quantity
\begin{eqnarray}
 \nonumber \dfrac{p \dot{V}}{V_0} &=& \dfrac{p}{V_0} \left[ C^{\text{eff}} \dot{X} + X \sum_{\alpha} \left( \dfrac{\partial S_C}{\partial \Lambda_{\alpha}} \right) \dot{\Lambda}_{\alpha} + \sum_{\alpha} \left( \dfrac{\partial V}{\partial \Lambda_{\alpha}} \right) \dot{\Lambda}_{\alpha} \right] \\ &\approx& p \epsilon_Z \left[ c_0 \left( \dot{\chi} + \dfrac{3 \dot{a}}{a} \chi \right) + \dfrac{2}{\chi^2} e^{-1/\chi} \dot{\chi} \right], 
\end{eqnarray}
the rate of work done against the confining pressure normalized by the extensive volume. It is negative because grain fragmentation produces small grains which fill the voids between large ones, causing overall volume compaction. The area shaded in cyan, enveloped by the curve $s \dot{\gamma}^{\text{pl}} / \dot{\gamma}$, then represents the heat dissipation per unit volume, being the difference between the total external plastic work -- attributed to shear deformation and compaction -- and the fracture energy. The dissipation rate to fracture goes asymptotically to zero at very large strains (of order $10$), implying that particle rearrangements dominate at large strains when the constituent grains are small enough. The downward cusp in the heat dissipation rate per unit strain at a shear strain of roughly $0.18$ in Fig.~\ref{fig:sxdispplotf51} for $\dot{\gamma} = 0.1$ s$^{-1}$, and at a shear strain of $0.27$ in Fig.~\ref{fig:sxdispplotf65} for $\dot{\gamma} = 10$ s$^{-1}$, indicates the abrupt emergence of grain-splitting events, as we shall see in Fig.~\ref{fig:axplot} below.

%%%%%%%%%%%%% FIGURE 5 %%%%%%%%%%%%
\begin{figure}[here]
\centering \includegraphics[width=.45\textwidth]{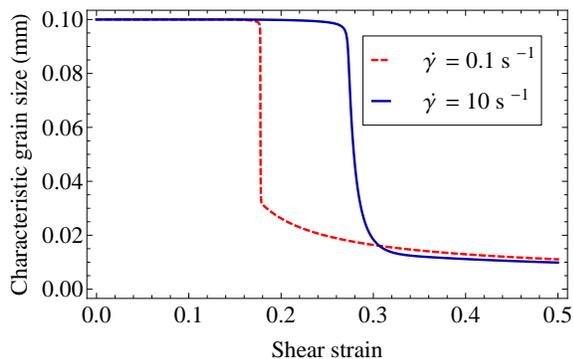} \caption{\label{fig:axplot}(Color online) Variation of typical grain size with accumulated shear strain, for two identical systems composed of grains that break apart, sheared at $\dot{\gamma} = 0.1$ s$^{-1}$ with $\chi(t = 0) = 0.055$ (red dashed curve), and at $\dot{\gamma} = 10$ s$^{-1}$ with $\chi(t = 0) = 0.065$ (dark blue solid curve), corresponding to the conditions in Figs.~\ref{fig:sxplotf51} and~\ref{fig:sxplotf65}, respectively. The fixed confining pressure is $p = 25$ MPa.} 
\end{figure}
%%%%%%%%%%%%%%%%%%%%%%%%%%%%%%%%%%%%%  

One can also examine the effect of variations in the imposed shear rate $\dot{\gamma}$ on grain size reduction. Figure~\ref{fig:axplot} shows the evolution of the typical grain size $a$ with the accumulated shear strain, corresponding to the two sets of initial conditions in Figs.~\ref{fig:sxplotf51} and \ref{fig:sxplotf65}, sheared at $\dot{\gamma} = 0.1$ s$^{-1}$ and $10$ s$^{-1}$, respectively. Indeed, as we have alluded to above, grain splitting -- the fragmentation of grains into many daughter grains whose size is only a fraction of the original, resulting in a drastic reduction of the characteristic grain size -- begins abruptly upon the onset of plastic deformation, at a shear strain of $0.18$ of $0.27$ in each of the two cases. At larger strains, however, grain size reduction occurs at at an appreciably slower rate. Our prediction is in accord with the hypothesis~\cite{sammis_1987} that most of the gouge material is formed at relatively low strain; it is also in conformity with the simulations of Mair and Abe~\cite{mair_2011}, who demonstrated that grain splitting is dominant in the initial stages of their simulations, while grain abrasion -- events for which the largest daughter grain and the original grain are of similar size -- become dominant at large shear displacement. The imposed shear rate appears to have no discernible effect on the long-time grain comminution behavior.

\section{Implication of grain breakage on strain localization in a granular material}

We now return to Figs.~\ref{fig:sxplotf51} and \ref{fig:sxplotf65} above, where we also showed the stress-strain response in the presence of strain localization, and shift our focus to the implication of grain comminution on localized deformation. Strain localization is commonplace in real granular materials such as fault gouge material; often, the shear band is comprised of finely comminuted particles that occupy a width of a few millimeters, 10 to 100 times a typical particle diameter~\cite{sammis_1987,mair_1999,rice_2006,chambon_2006}. Shear banding was previously exhibited in amorphous, molecular solids within the STZ framework~\cite{manning_2007a,manning_2009,daub_2009}; there is no analog of grain comminution in molecular glasses, and a model of strain localization in granular materials ought to be able to predict the inception of a finely granulated layer upon the onset of shear deformation. This section investigates the implication of grain breakage and strain localization on granular flow rheology.

To this end, we include, as in molecular glasses~\cite{manning_2007a,manning_2009,daub_2009} a diffusion term in the ``heat equation'' Eq.~\eqref{eq:Xdot2} for the compactivity. That equation now becomes (we set $\theta = 0$ as before for athermal situations)
\begin{equation}\label{eq:Xdot3}
 C^{\text{eff}} \dot{X} = V \dfrac{s}{p} \dot{\gamma}^{\text{pl}} - {\cal K} X + V_0 \dfrac{\gamma_G}{p a^2}\dot{a} + C^{\text{eff}} \dot{\gamma}^{\text{pl}} \dfrac{\partial}{\partial y} \left( D_0 a^2 \dfrac{\partial X}{\partial y} \right),
\end{equation}
where $y$ is the position coordinate along the width of the sheared material. $D_0 a^2$ is the analog of the ``effective heat diffusion'' coefficient in earlier work; we have explicitly written it as the product of $a^2$ and the dimensionless number $D_0$ to emphasize that the diffusion length scale is set by the characteristic grain size. Reverting to the dimensionless compactivity $\chi$, its temporal evolution equation reads
\begin{eqnarray}\label{eq:chidot2}
 \nonumber \dot{\chi} &=& \left( 1 - \dfrac{\chi}{\hat{\chi} (q)} \right) \left( \dfrac{s}{p} \right) \dot{\gamma}^{\text{pl}} \left( \dfrac{1 - \kappa_0 e^{- \gamma_G / p a}}{c_0 \epsilon_Z} \right) \\ & & + \dfrac{D_0 \dot{\gamma}^{\text{pl}}}{a^3} \dfrac{\partial}{\partial y} \left[ a^2 \dfrac{\partial}{\partial y} \left( a^3 \chi \right) \right] .
\end{eqnarray}
The shear stress $s$ is assumed to equilibrate rapidly across the sample and is therefore assumed to be uniform; its governing equation, Eq.~\eqref{eq:sdot} becomes
\begin{equation}\label{eq:sdot2}
 \dot{s} = G \left( \dot{\gamma} - \dfrac{1}{2 L} \int_{-L}^L \dot{\gamma}^{\text{pl}} dy \right) ,
\end{equation}
and the temporal evolution equation for characteristic grain size $a$ at each position $y$, Eq.~\eqref{eq:adot1}, remains unchanged:
\begin{equation}\label{eq:adot3}
 \dot{a} = - \kappa_0 \exp \left( - \dfrac{p_{\text{th}}}{p} \right) \dfrac{\dot{\gamma}^{\text{pl}} s}{\gamma_G} a^2 .
\end{equation}
As our initial condition, we assume that the grain size distribution is uniform, $a(y, t = 0) = a_0 = \text{constant}$, but that the there is slightly more disorder towards the middle of the sample:
\begin{equation}\label{eq:x_init}
 \chi (y, t = 0) = \bar{\chi} + \chi_p \sech \left( \dfrac{k y}{L} \right).
\end{equation}
for $- L \leq y \leq L$, where $k$ is a dimensionless number; disorder is therefore spread over a width of $L / k$. This type of initial condition has been found to result in the formation of shear bands in the case of molecular glasses~\cite{manning_2007a,manning_2009,daub_2009}.

In this section, we demonstrate several possible scenarios that can arise from different initial conditions, in order to understand the implication of grain fragmentation on shear localization and granular flow rheology. To this end, the STZ dynamical equations, Eqs.~\eqref{eq:chidot2},~\eqref{eq:sdot2} and~\eqref{eq:adot3}, are solved numerically in an irregular spatial mesh to resolve shear localization; we apply central differencing to the spatial derivatives and use an implicit time-stepping scheme. Unless otherwise specified, we employ the same parameters as in Sec.~VIII and use, in addition, $D_0 = 1$ for the diffusion coefficient and $L = 1$ m for the half-width of the material. We set the spread of the initial perturbation by choosing $k = 10$; we only vary the amount of disorder through the parameters $\bar{\chi}$ and $\chi_p$ in Eq.~\eqref{eq:x_init}, and vary the imposed strain rate $\dot{\gamma}$.

\subsection{Disorder-limited shear bands}

If the granular material is sufficiently aged, and sheared slowly enough, one observes violent fragmentation within the shear band that occupies a finite width within the material, the hallmark of disorder-limited strain localization~\cite{manning_2009}. In this subsection, we use $\bar{\chi} = 0.055$ for the initial disorder, with $\chi_p = \bar{\chi}/8$ and $0$ for strain localization and homogeneous deformation, respectively. In addition, we impose the fixed total strain rate $\dot{\gamma} = 0.1$ s$^{-1}$.

%%%%%%%%%%%%% FIGURE 6 %%%%%%%%%%%%
\begin{figure}[here]
\centering
% \subfigure{\includegraphics[width=.45\textwidth]{fig_aytplot3df51s.eps}\label{fig:ayplotf51a}}
% \subfigure{\includegraphics[width=.45\textwidth]{fig_ayplotf51b.eps}\label{fig:ayplotf51b}}
\includegraphics[width=.45\textwidth]{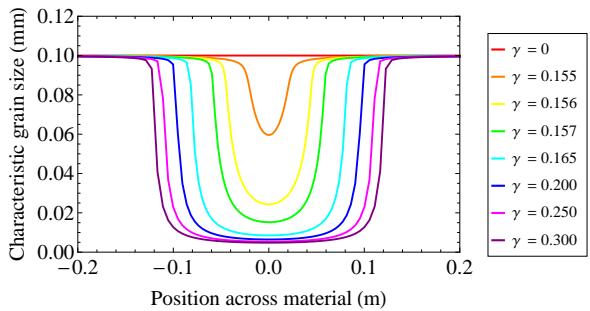} \caption{\label{fig:ayplotf51}(Color online) Variation of characteristic grain size $a$ across the width of the material, at various accumulated shear strains $\gamma \equiv \dot{\gamma} t$, for a disorder-limited shear band. Grains in the shear band are comminuted more finely than those away from the shear band to sizes of order $5$ $\mu$m. The shear band saturates at about $10\%$ the width of the material. For clarity, we show only the grain size evolution within the middle $20\%$ material width.} 
\end{figure}
%%%%%%%%%%%%%%%%%%%%%%%%%%%%%%%%%%%%% 

%%%%%%%%%%%%% FIGURE 5 %%%%%%%%%%%%
\begin{figure}[t]
\centering 
\subfigure{\includegraphics[width=.45\textwidth]{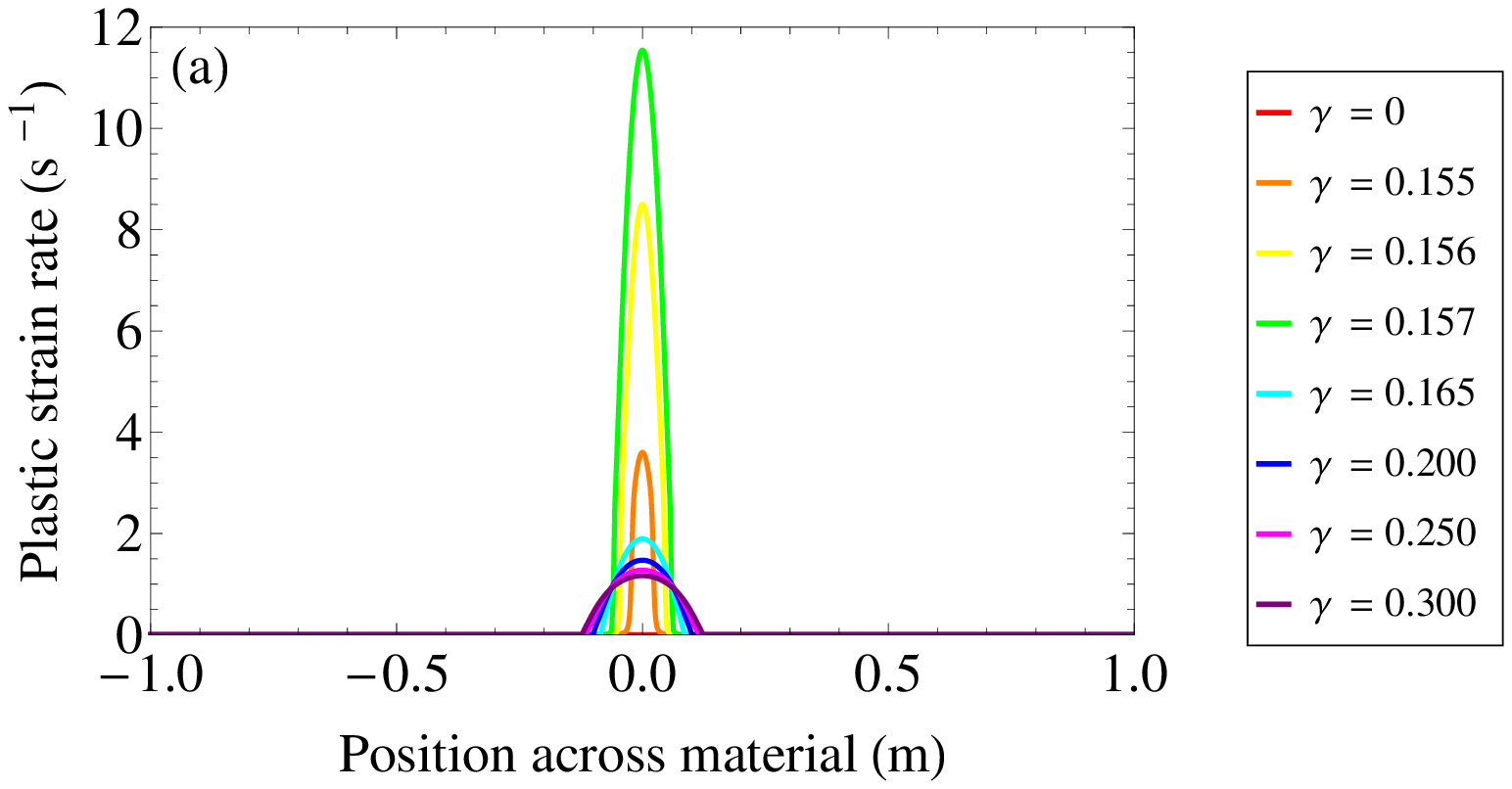}\label{fig:vyplotf51a}}
\subfigure{\includegraphics[width=.45\textwidth]{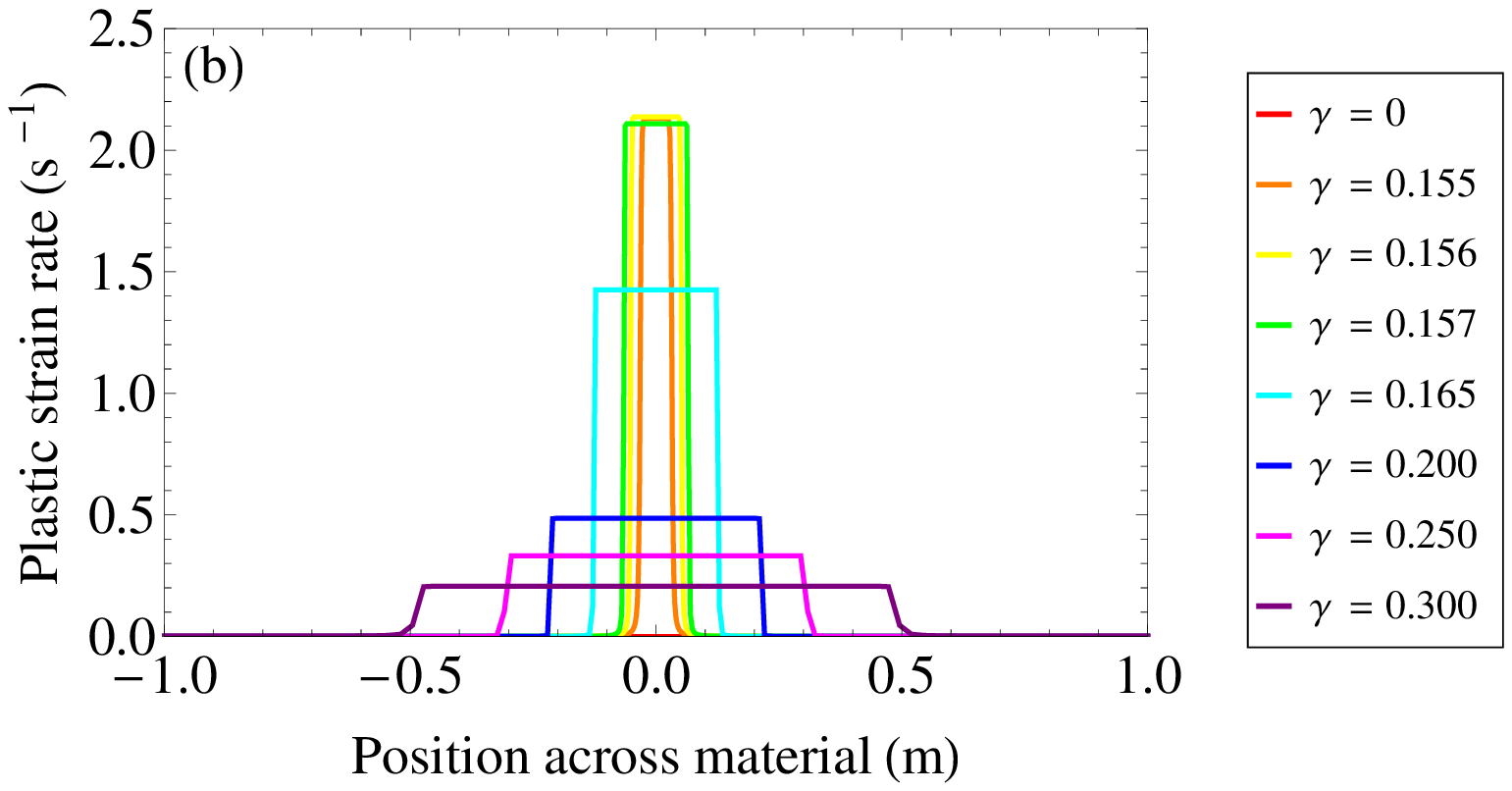}\label{fig:vyplotf51b}}
\caption{\label{fig:vyplotf51}(Color online) Variation of plastic strain rate $\dot{\gamma}^{\text{pl}}$ across the width of the material, at various accumulated shear strains $\gamma \equiv \dot{\gamma} t$, for a disorder-limited shear band, in the (a) presence and (b) absence of grain fragmentation. The applied strain rate is $\dot{\gamma} = 0.1$ s$^{-1}$. Shortly after the onset of plastic deformation at $\gamma \sim 0.15$, the localized disorder about $y = 0$ causes intense comminution within the shear band, when grain fragmentation does operate. In such a case, the shear band appears to saturate at about $10\%$ the width of the material. When grain fragmentation does not operate, however, the shear band gradually diffuses outward.} 
\end{figure}
%%%%%%%%%%%%%%%%%%%%%%%%%%%%%%%%%%%%% 

We refer to Fig.~\ref{fig:sxplotf51} in the last section for the stress-strain response; there, we have shown that grain breakage causes the shear stress to approach the steady-state value faster when the shear strain is localized; grain breakage is therefore a softening mechanism in the presence of strain localization. Meanwhile, Fig.~\ref{fig:ayplotf51} shows that grains in the shear band are subject to intense fragmentation, while those away from the shear band are not affected, in conformity with experimental and field observations. At a shear strain of $\gamma = 0.3$, the characteristic grain size at the middle of the shear band roughly equals $5 \mu$m, which is $ 5\%$ the grain size at the edge of the material.

\begin{figure}[here]
\centering 
\subfigure{\includegraphics[width=.45\textwidth]{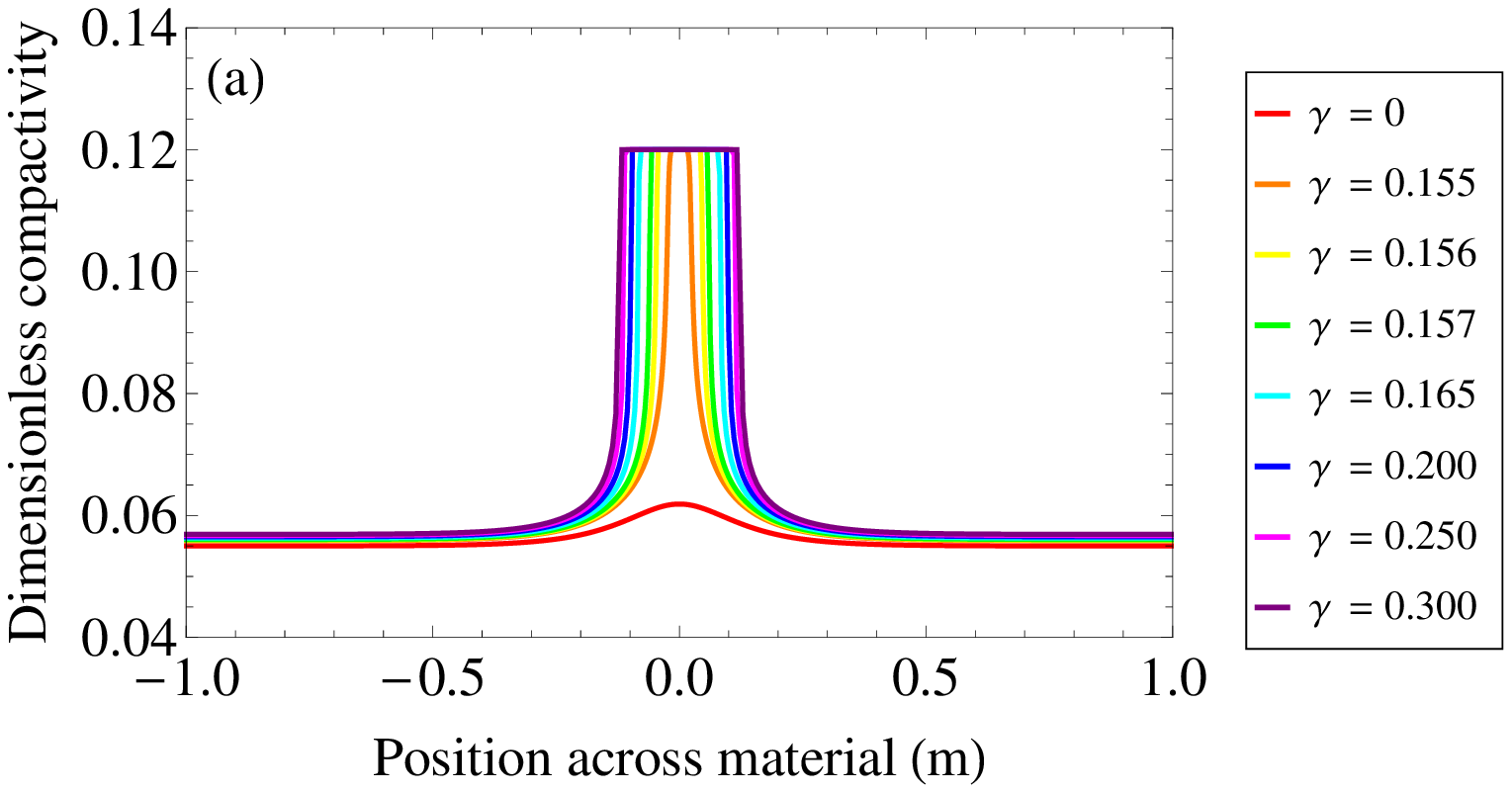}\label{fig:xyplotf51a}}
\subfigure{\includegraphics[width=.45\textwidth]{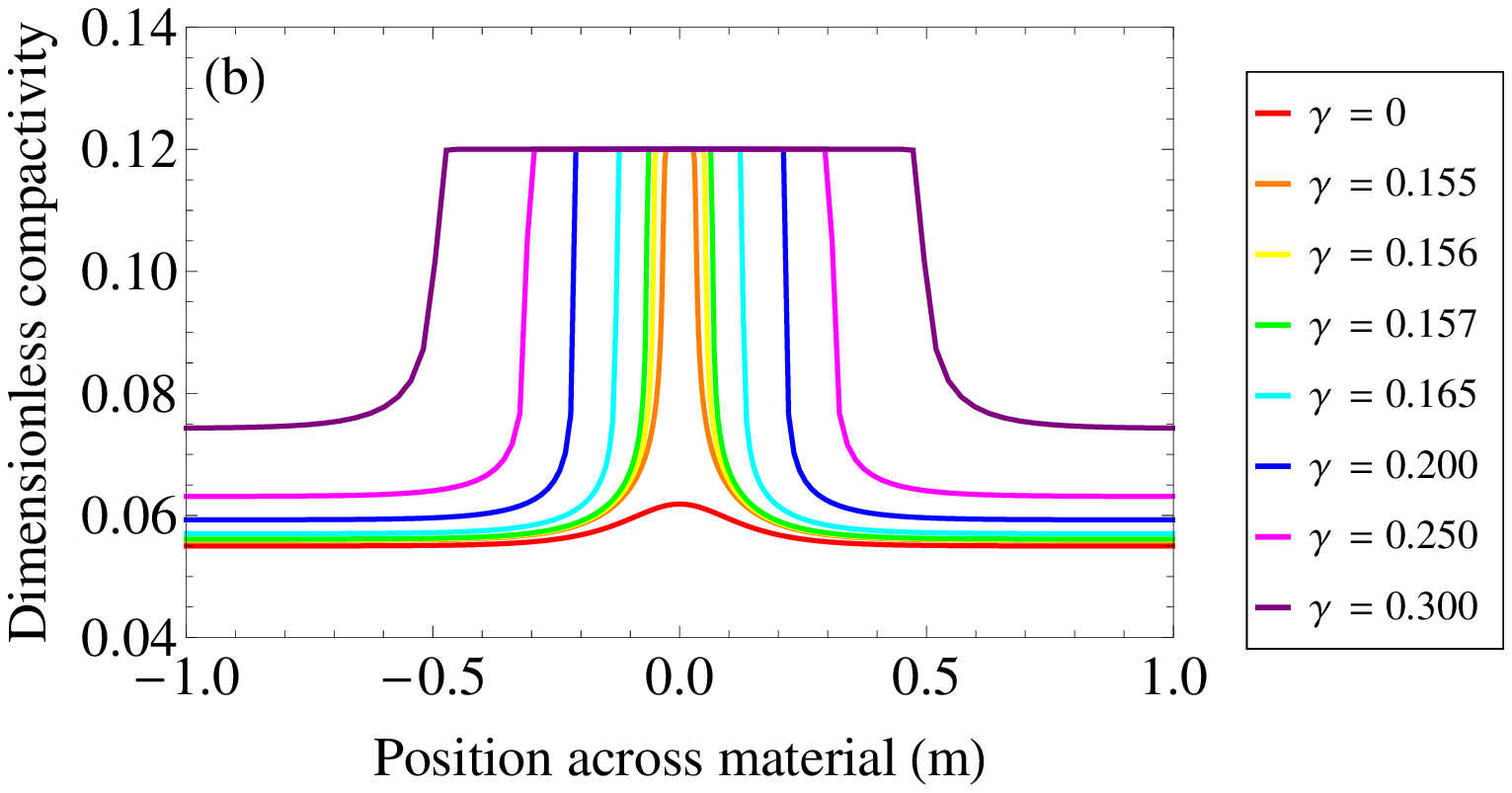}\label{fig:xyplotf51b}}
\caption{\label{fig:xyplotf51}(Color online) Variation of dimensionless compactivity $\chi$ across the width of the material, as a function of accumulated shear strain $\gamma = \dot{\gamma} t$, for a disorder-limited shear band, in the (a) presence and (b) absence of grain fragmentation. Within the comminuted shear band, $\chi = \hat{\chi}_0$; away from it, there is no plastic deformation and $\chi$ never approaches $\hat{\chi}_0$.} 
\end{figure}

Figures~\ref{fig:vyplotf51} and~\ref{fig:xyplotf51} show the effect of grain breakage on the distribution of plastic strain rate and configurational disorder, when strain localization occurs. Upon the formation of the shear band, the relaxation of strain localization inside the material is significantly slowed down by grain fragmentation, as depicted by Fig.~\ref{fig:vyplotf51a}; for our choice of parameters, the shear band appears to saturate at roughly $10\%$ the width of the material. In comparison, the shear band gradually diffuses outward if grains do not break apart, as seen in Fig.~\ref{fig:vyplotf51b}. In both cases the dimensionless compactivity $\chi$ becomes rather uniform across the shear band shortly after the onset of plastic deformation: $\chi \approx \hat{\chi}_0$ across the comminuted region. Once again, Fig.~\ref{fig:xyplotf51a} shows that when grain fragmentation occurs, the shear band becomes essentially frozen, with no discernible diffusion of configurational disorder, in contrast to Fig.~\ref{fig:xyplotf51b} which shows that configurational disorder slowly diffuses across the material in the absence grain breakage.

%%%%%%%%%%%%% FIGURE 7 %%%%%%%%%%%%
%%%%%%%%%%%%%%%%%%%%%%%%%%%%%%%%%%%%% 

We propose to explain these observations in terms of the lubrication effect that small particles have on the rolling motion of large particles~\cite{reches_2010,han_2011}. The initial condition for $\chi (t = 0)$ is inhomogeneous; there is a spike at the middle axis, $y = 0$, of the material. This implies that upon the onset of plastic deformation, the plastic strain rate $\dot{\gamma}^{\text{pl}}$ is larger at $y = 0$. Since the rate of grain size reduction, Eq.~\eqref{eq:adot3}, is proportional to the amount of plastic work per unit volume, this heterogeneity enhances grain fragmentation about the center of the material. The smaller grains produced at the center lubricate the surrounding grains, causing the diffusion of configurational disorder via the diffusion term in Eq.~\eqref{eq:Xdot3} and promoting slip motion. The corresponding increase in the plastic strain rate nearby intensifies grain fragmentation and causes the spread of the width of the comminuted region. However, the diffusion length scale, proportional to $a^2$, sharply decreases as a result of grain size reduction. In addition, the sharp grain size gradient at the interface between the comminuted and uncomminuted regions provides lubrication and shields the large grains from fracturing. These effects slow down the diffusion of configurational disorder across the entire width of the material.

Said differently, the large grains just outside of the shear band, now lubricated by the fine grains at the edge of the shear band, move in concert with the uncomminuted grains far away from the shear band. The plastic strain rate in the uncomminuted region is therefore zero; since the diffusion coefficient in Eq.~\eqref{eq:Xdot3} is proportional to the strain rate, diffusion of disorder stops. From this point on, plastic deformation and grain fragmentation is confined within the shear band; the shear band diffuses very slowly outward, if at all. Configurational disorder is now uniformly distributed within the shear band with $\chi \approx \hat{\chi}_0$ (Fig.~\ref{fig:xyplotf51}), but particles at the middle of the shear band, $y = 0$, are smaller than those nearby (Fig.~\ref{fig:ayplotf51}). According to Eq.~\eqref{eq:strainrate} and the remarks that follow, the STZ time scale $\tau$ is proportional to the characteristic grain size $a$; thus, with the same amount of configurational disorder, the plastic strain rate $\dot{\gamma}^{\text{pl}}$ increases with decreasing grain size. This accounts for a higher plastic strain rate at the middle of the shear band than closer to its edge. We noted above that at a shear strain of $\gamma = 0.3$, the grain size at the center of the shear band equals 5 $\mu$m, which is below the grind limit given above in Sec.~V. Therefore, grain comminution slows down appreciably, granular rearrangement becomes the dominant dissipative mechanism, and the shear band appears to be long-lived.

While not shown here, the width of the finely comminuted shear band decreases upon decreasing $\chi(t = 0)$, the initial amount of configurational disorder (or configurational dilatation), within the disorder-limited shear localization regime.

\subsection{Diffusion-limited shear bands}

There is yet another class of shear localization phenomena for which there is no well-defined boundary between the intensely comminuted region and the surrounding material; the entire granular system experiences grain fragmentation, though the extent of fragmentation depends on the distance from the central shear band. The shear band, whose width is a function of the initial conditions, persists at large shear strains; relaxtion to homogeneous deformation occurs extremely slowly. In addition, the dimensionless compactivity $\chi$ equilibrates to $\hat{\chi}_0$, the steady-state value, across the entire material. This kind of scenario arises primarily when the initial compactivity is somewhat large, and occasionally when the shear rate is sufficiently fast.

In this subsection, we start with a material with identical initial shear stress, confining pressure and grain size as the one in Subsection A in which disorder-limited shear localization occurred -- i.e.,~$s(t = 0) = 0.01$ Pa, $p = 25$ MPa, and $a(t = 0) = 0.1$ mm -- but with less aging and more configurational disorder. Thus $\chi(t = 0) = \bar{\chi} + \chi_p \sech (k y / L)$ with $\bar{\chi} = 0.065$, $\chi_p = \bar{\chi} / 8$, and $k = 10$. In addition, we impose a higher strain rate $\dot{\gamma} = 10$ s$^{-1}$. We explore the implication of grain fragmentation on flow rheology at this large shear rate; we also examine the effect of strain localization by numerically integrating the STZ equations of motion, Eqs.~\eqref{eq:chidot2},~\eqref{eq:sdot2} and ~\eqref{eq:adot3}, in the case of homogeneous flow, where we set $\chi_p = 0$, and compare that to the spatially heterogeneous solution. Figure~\ref{fig:sxplotf65} in the previous section shows that at this fast shear rate, grain fragmentation significantly softens the material, independent of strain localization. This validates the observation by Han et al.~\cite{han_2011} that fine particles lubricate gouge material and result in significant flow stress reduction at fast shear rates. The same figure also shows that strain localization provides another weakening mechanism by lowering the peak stress.

%%%%%%%%%%%%% FIGURE 8 %%%%%%%%%%%%
\begin{figure}[here]
\centering 
\includegraphics[width=.45\textwidth]{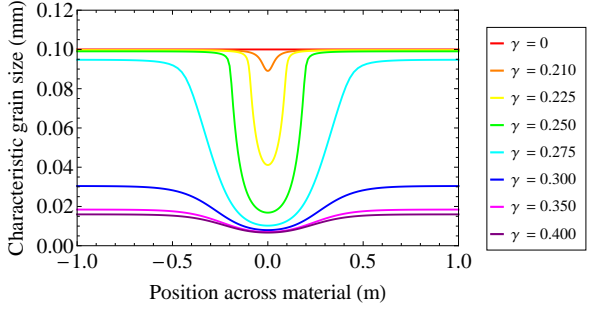} \caption{\label{fig:ayplotf65}(Color online) Variation of characteristic grain size $a$ across the width of the material, at various accumulated shear strains $\gamma = \dot{\gamma} t$, for a diffusion-limited shear band. The perturbation to the initial configurational disorder does result in the formation of a comminuted zone whose width is about $1/4$ of the whole material, but the surrounding region also experiences continuous grain fragmentation, though to a somewhat lesser extent. At a shear strain of $\gamma = 0.40$, the grain size at the middle of the shear band and at the edge of the material equal 6.5 $\mu$m and 16 $\mu$m, respectively.} 
\end{figure}
%%%%%%%%%%%%%%%%%%%%%%%%%%%%%%%%%%%%% 

\begin{figure}[here]
\centering 
\subfigure{\includegraphics[width=.45\textwidth]{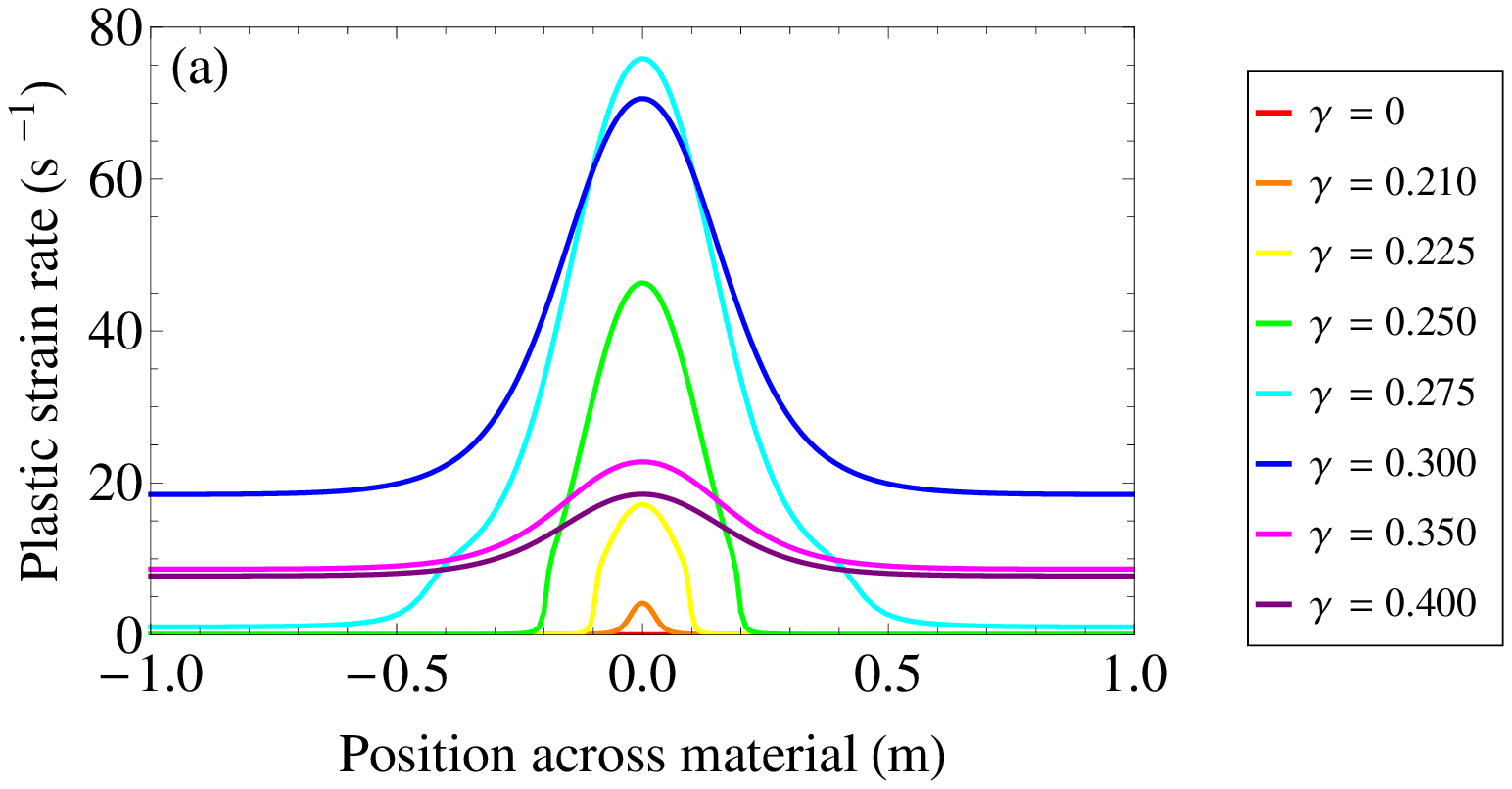}\label{fig:vyplotf65a}}
\subfigure{\includegraphics[width=.45\textwidth]{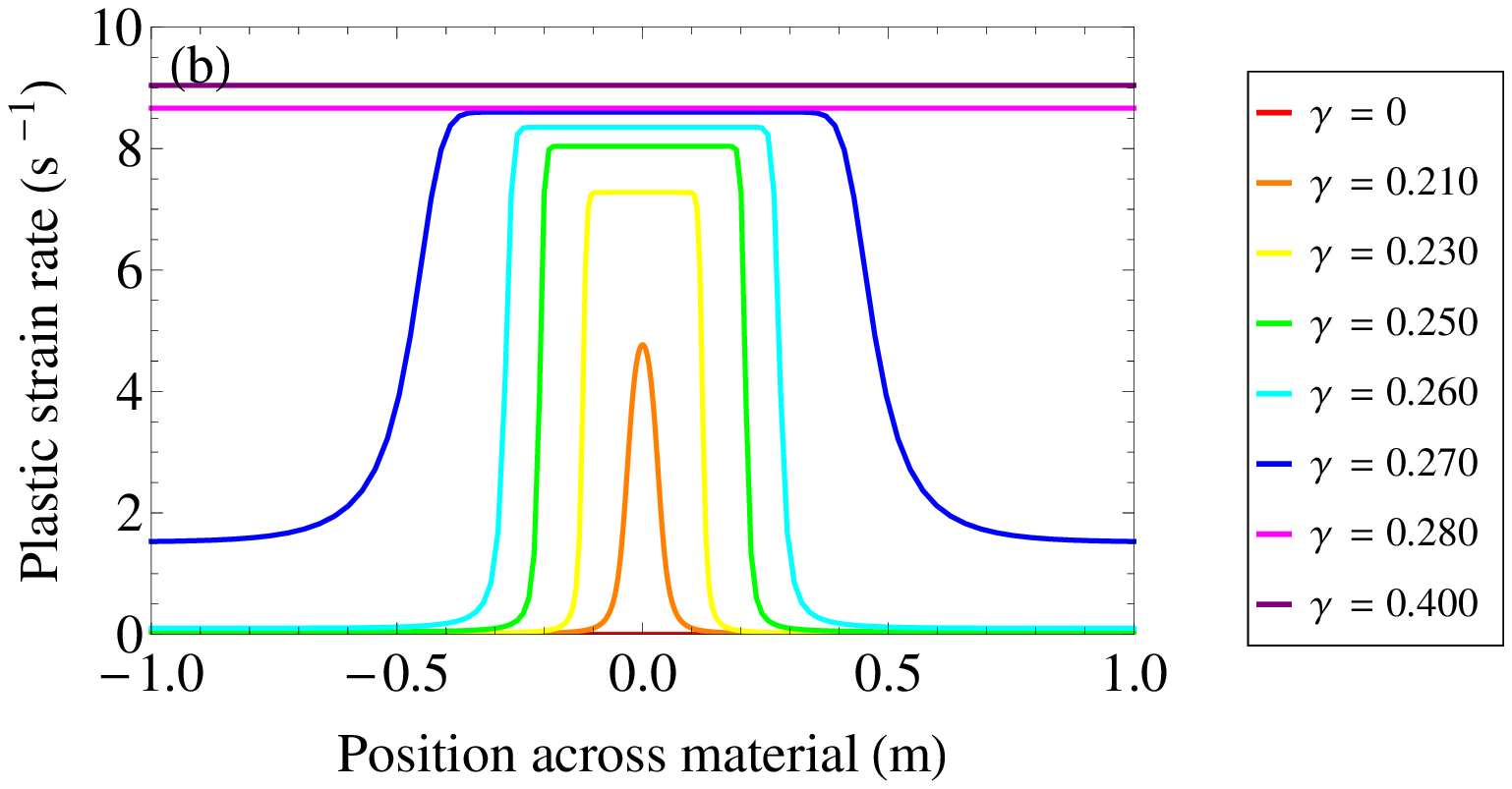}\label{fig:vyplotf65b}}
\caption{\label{fig:vyplotf65}(Color online) Variation of plastic strain rate $\dot{\gamma}^{\text{pl}}$ across the width of the material, at various accumulated shear strains $\gamma = \dot{\gamma} t$, for a diffusion-limited shear band, in the (a) presence and (b) absence of grain fragmentation. The imposed strain rate is $\dot{\gamma} = 10$ s$^{-1}$. When subjected to grain breakage, the plastic strain rate within the shear band rapidly increases upon its nucleation, and peaks at around $\gamma \simeq 0.25$, before relaxing towards a long-lived quasi-static state with a peak strain rate of $\dot{\gamma}^{\text{pl}} \simeq 17$ s$^{-1}$ at the center, $70\%$ above the imposed strain rate. In contrast, the initial heterogeneity quickly relaxes to produce a spatially uniform strain rate profile when grain breakage does not occur.} 
\end{figure}

The temporal evolution of the characteristic grain size $a$ and the plastic strain rate $\dot{\gamma}^{\text{pl}}$, shown in Figs.~\ref{fig:ayplotf65} and~\ref{fig:vyplotf65a} respectively, exhibit important features qualitatively different from a disorder-limited shear band above. While significant grain size reduction and accumulation of plastic strain occur near the center of the material, a sharp boundary between the comminuted region and the surrounding region is not observed; instead, particles across the entire material experience fragmentation. Moreover, the plastic strain rate across the material appears to stabilize shortly after reaching the yield stress at $\gamma \simeq 0.27$, and grain fragmentation near the center of the shear band retards significantly thereafter. Direct comparison between Figs.~\ref{fig:vyplotf65a} and~\ref{fig:vyplotf65b} shows that comminution promotes shear localization, as in a disorder-limited shear band above (Fig.~\ref{fig:vyplotf51}); in fact, in the present case, the effect of comminution on shear banding is even more pronounced. The plastic strain rate distribution remains nonuniform in the presence of grain breakage, but when grain fragmentation does not occur, deformation becomes spatially uniform.

%%%%%%%%%%%%% FIGURE 9 %%%%%%%%%%%%
%%%%%%%%%%%%%%%%%%%%%%%%%%%%%%%%%%%%%

\begin{figure}[here]
\centering \includegraphics[width=.45\textwidth]{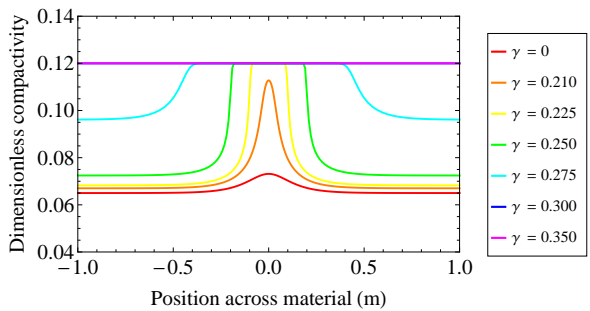} \caption{\label{fig:xyplotf65}(Color online) Variation of dimensionless compactivity $\chi$ across the width of the material, at various accumulated shear strains $\gamma = \dot{\gamma} t$, for a diffusion-limited shear band subject to grain fragmentation. At large strains, $\chi \rightarrow \hat{\chi}_0$ across the entire material. The temporal evolution of compactivity without grain breakage is quantitatively similar and is therefore not separately shown.} 
\end{figure}

The larger initial compactivity $\chi(t = 0)$ in here is key to explaining these qualitative differences that distinguish a diffusion-limited shear band from the disorder-limited shear band in Subsection A. (Indeed, at $\bar{\chi} = 0.065$, these qualitative features persist down to an imposed strain rate of $\dot{\gamma} = 0.1$ s$^{-1}$; we used $\dot{\gamma} = 10$ s$^{-1}$ here only to depict the softening effect of grain fragmentation at fast shear rates, independent of strain localization.) To see this, remember that the STZ density $\Lambda = 2 e^{-1 / \chi}$ upon the onset of plastic deformation increases rapidly with $\chi(t = 0)$, and that the plastic strain rate $\dot{\gamma}^{\text{pl}}$ is proportional to $\Lambda$. Because the diffusion term in Eq.~\eqref{eq:Xdot3} is proportional to $\dot{\gamma}^{\text{pl}}$, the increased plastic strain rate facilitates diffusion of configurational disorder, and therefore the spread of plastic strain, throughout the entire material. This enables the dimensionless compactivity $\chi$ to equilibrate to $\hat{\chi}_0$ across the material (Fig.~\ref{fig:xyplotf65}). The equilibration of $\chi$ to $\hat{\chi}_0$, however, occurs while plastic strain remains localized in the shear band. Now that $\dot{\chi} \simeq 0$ (because $\chi \simeq \hat{\chi}_0$ and the diffusion term vanishes), the localized plastic shear becomes long-lived. Relaxtion of the localized strain rate distribution occurs very slowly. The large imposed shear rate, and the absence of a sharp boundary that differentiates the comminuted region from the surrounding particles, permit the dissipation of energy throughout the material.

%%%%%%%%%%%%% FIGURE 10 %%%%%%%%%%%%
%%%%%%%%%%%%%%%%%%%%%%%%%%%%%%%%%%%%% 

It is interesting to note that with $\tau = a \sqrt{ \rho_G / p}$ (see the remark following Eq.~\eqref{eq:strainrate}), the extremely fine grains at the middle of the shear band in fact correspond to a smaller dimensionless plastic strain rate $q = \tau \dot{\gamma}^{\text{pl}}$, for a given $\dot{\gamma}$. Thus, for the diffusion-limited shear band above in Fig.~\ref{fig:vyplotf65a}, a strain rate of $\dot{\gamma}^{\text{pl}} \approx 18.5$ s$^{-1}$ at the middle of the shear band, where the grain size approaches $a \approx 6.5 \mu$m at a shear strain of $0.4$, corresponds to a dimensionless plastic strain rate of $q = 9.62 \times 10^{-7}$, which is an order of magnitude smaller than the dimensionless strain rate of $q = 8 \times 10^{-6}$ when grains do not break apart.

% With hindsight, the same argument also justifies self-consistently our prior assumption that $\hat{\chi} (q) \approx \hat{\chi}_0$ for the steady-state dimensionless compactivity at strain rate $q$, even though in units of inverse time, $\dot{\gamma}^{\text{pl}}$ within the shear band may be a few orders of magnitude larger than the imposed strain rate $\dot{\gamma}$. For example, fine grains of size $a \approx 5 \times 10^{-6}$ m at the middle of the disorder-limited shear band, where the local plastic strain rate equals $\dot{\gamma}^{\text{pl}} \approx 1.17$ s$^{-1}$ in Fig.~\ref{fig:vyplotf51a}, corresponds to $q \approx 4 \times 10^{-8}$, small enough for the $\hat{\chi}(q) \approx \hat{\chi}_0$ approximation to hold extremely well.

\subsection{Summary: Localization behavior as a function of initial preparation and strain rate}

In summary, if the initial compactivity $\chi (t = 0)$ is large enough (i.e.~if the granular material is loose) and is spatially heterogeneous, the resultant shear band is of the diffusion-limited type. In contrast, if $\chi (t = 0)$ is small enough (i.e.~if the material has undergone a large amount of aging), and if the shear rate is slow, the shear band is disorder-limited, with a sharp boundary between the comminuted and uncomminuted regions.

\section{Concluding remarks}

In this paper we have developed a model for grain fragmentation processes within the framework of nonequilibrium thermodynamics. By incorporating the model into the STZ theory for hard-core materials, where the characteristic grain size $a$ now becomes a dynamical variable, we have analyzed the implications of grain breakage on the frictional and shear deformation behavior of granular materials. We have found that grain breakage serves as a weakening mechanism when shear localization occurs, and most remarkably, when the strain rate $\dot{\gamma}$ is fast if the material is rate-strengthening. In the latter case, grain breakage results in a sharp drop in the flow stress, and dissipates energy through the creation of new grain surfaces. Grain size reduction is most abrupt shortly after the onset of plastic strain, and slows down considerably afterwards, indicating that grain splitting is dominant at small strains while grain abrasion becomes prominent at large strains.

In addition, we have shown that there exists a feedback between disorder configuration and grain fragmentation; a variety of distinct qualitative shear localization behaviors, dependent on the initial state of the granular material, may occur. In one type of localization -- the disorder-limited shear band -- there is a clear boundary between the finely comminuted region and surrounding particles. This type of behavior occurs for a sufficiently aged, or densely packed, granular material. The metastable boundary between the comminuted region and the surrounding particles is associated with the lubrication effect that small, nano-sized particles exert on larger particles nearby, as seen in model fault gouge~\cite{reches_2010,han_2011}.

Another type of localization -- the diffusion-limited shear band -- occurs in looser granular packings. There is no sharp boundary that separates the comminuted region from the rest of the material; all particles are crushed, though to different extents at different positions. This is accounted for by the augmented STZ density that increases the plastic strain rate $\dot{\gamma}^{\text{pl}}$ drives the diffusion of configurational disorder across the material. One remarkable feature is that the dimensionless compactivity $\chi$ equilibrates across the material to its steady-state value $\hat{\chi}_0$ before plastic strain becomes delocalized, leading to a long-lived quasi-static state of inhomogeneous deformation; it relaxes to homogeneous deformation very slowly, if at all.

Geophysical observations indicate that in a sheared fault gouge, particles within the shear band are significantly smaller than those outside of the shear band~\cite{sammis_1987,mair_1999,rice_2006,chambon_2006}. This appears to validate our large-strain predictions. However, to the best of our knowledge, there appear to be no direct verification of the short-time dynamical behavior, nor is there a direct verification of our proposed grain size evolution equation, Eq.~\eqref{eq:adot1}, which is based solely on physical grounds. Because of this, it is not possible to rigorously constrain the various parameters in our theory; in this study, some of them have been chosen based on knowledge of the material properties of sand particles, while other parameter values have been chosen to display the distinctive qualitative implications that grain breakage have on the dynamical material behavior. Do similar qualitative features arise if a different evolution equation for the grain size $a$ is used? It would be beneficial if future experiments and simulations could direct track the temporal evolution of the characteristic grain size, and validate Eq.~\eqref{eq:adot1} and its various ingredients.

With the simplification that ${\cal C}(s) = R_0 = \text{constant}$ and $s_0 = \text{constant}$, we implicitly assumed that inter-particle friction, grain angularity, and the effects of grain rotation versus slipping, etc., are subsumed under the choice of parameters, and that the rate and forces associated with granular rearrangement is independent of whether it occurs by slipping or rotation. This approach is certainly justified if grain fragmentation has only minor effects on grain angularity and roughness, as in the experiments on angular grains by Mair, Frye and Marone~\cite{mair_2002}, in which grain breakage is neither intrinsically strengthening nor weakening. Other experiments, however, point to the effect of grain characteristics on flow rheology; fragmentation of spherical grains into angular ones could strengthen the material~\cite{mair_2002}, while abrasion could result in increase in roundedness and provide weakening effects at high shearing velocities~\cite{han_2011}. As such, the microscopic physics of friction, angularity, sliding and rotation could have important implications on elements of the model and, as a result, the macroscopic behavior. We speculate that a multi-species STZ model~\cite{langer_2011b,langer_2012}, with a system-specific distribution of activation stresses and transition rates corresponding to different grain characteristics and deformation mechanisms, could provide a description of the second-order effects of grain characteristics on macroscopic behavior. The effect of grain angularity and friction on the transition rate ${\cal R}(s)$ and the activation stress $s_0$ would be the first and foremost issue that needs to be addressed in a refined model.

In this paper we employed the simplifying assumption that STZ transitions and grain breakage are independent of thermal temperature. In a granular fault gouge, however, various thermal processes (such as flash heating~\cite{rice_2006}) become prominent; they may have important implications on material properties. Would this influence the STZ transition rate ${\cal R} (s)$ or the rate of grain size reduction, or would they cause broken-up particles to recombine? The modeling of such processes would be an interesting avenue for further research.

~~~~~

\section*{Acknowledgments}

We thank James Langer for illuminating discussions. We also thank the anonymous reviewer whose well-informed comments helped significant improve the clarity of this manuscript. This work was supported by an Office of Naval Research MURI Grant No. N000140810747, NSF Grant No. DMR0606092, and the NSF/USGS Southern
California Earthquake Center, funded by NSF Cooperative Agreement EAR-0529922 and USGS Cooperative Agreement 07HQAG0008, and the David and Lucile Packard Foundation.


\begin{thebibliography}{5}

\bibitem{lieou_2012}C. K. C. Lieou and J. S. Langer, Phys. Rev. E \textbf{85}, 061308 (2012).

\bibitem{falk_1998}M. L. Falk and J. S. Langer, Phys. Rev. E \textbf{57}, 7192 (1998).

\bibitem{langer_2011}M. L. Falk and J. S. Langer, Ann. Rev. Cond. Matt. Phys. \textbf{2}, 353 (2011).

\bibitem{manning_2007a}M. L. Manning, J. S. Langer and J. M. Carlson, Phys. Rev. E \textbf{76}, 056106 (2007).

\bibitem{langer_2008}J. S. Langer, Phys. Rev. E \textbf{77}, 021502 (2008).

\bibitem{langer_2011a}E. Bouchbinder and J. S. Langer, Phys. Rev. Lett. \textbf{106}, 148301 (2011).

\bibitem{langer_2011b}E. Bouchbinder and J. S. Langer, Phys. Rev. E \textbf{83}, 061503 (2011).

\bibitem{langer_2012}J. S. Langer, Phys. Rev. E \textbf{85}, 051507 (2012).

\bibitem{edwards_1989a}S. F. Edwards and R. B. S. Oakeshott, Physica A \textbf{157}, 1080 (1989).

\bibitem{edwards_1989b}A. Mehta and S. F. Edwards, Physica A \textbf{157}, 1091 (1989).

\bibitem{edwards_1989c}S. F. Edwards and R. B. S. Oakeshott, Physica D \textbf{38}, 88 (1989).

\bibitem{edwards_1990a}S. F. Edwards, Rheol. Acta \textbf{29}, 493 (1990).

\bibitem{edwards_1990b}S. F. Edwards, J. Phys. Condensed Matter \textbf{2}, SA63 (1990).

\bibitem{liu_2011}T. K. Haxton, M. Schmiedeberg and A. J. Liu, Phys. Rev. E \textbf{83}, 031503 (2011).

\bibitem{haxton_2012}T. K. Haxton, Phys. Rev. E \textbf{85}, 011503 (2012).

\bibitem{mair_2011}K. Mair and S. Abe, Pure. Appl. Geophys. \textbf{168}, 2277 (2011).

\bibitem{sammis_1987}C. Sammis, G. King, and R. Biegel, Pure. Appl. Geophys. \textbf{125}, 777 (1987).

\bibitem{reches_2010}Z. Reches and D. A. Locker, Nature (London) \textbf{467}, 452 (2010).

\bibitem{han_2011}R. Han, T. Hirose, T. Shimamoto, Y. Lee, and J. Ando, Geology \textbf{39}, 599 (2011).

\bibitem{mair_2002}K. Mair, K. M. Frye, and C. Marone, J. Geophys. Res. \textbf{107} (B10), 2219 (2002).

\bibitem{guo_2006}Y. Guo and J. K. Morgan, J. Geophys. Res. \textbf{111}, B12406 (2006).

\bibitem{mair_1999}K. Mair and C. Marone, J. Geophys. Res. \textbf{104}, 28899 (1999).

\bibitem{rice_2006}J. R. Rice, J. Geophys. Res. \textbf{111}, B05311 (2006).

\bibitem{chambon_2006}G. Chambon, J. Schmittbuhl, and A. Corfdir, J. Geophys. Res. \textbf{111}, B09308 (2006).

\bibitem{manning_2009}M. L. Manning, E. G. Daub, J. S. Langer, and J. M. Carlson, Phys. Rev. E \textbf{79}, 016110 (2009).

\bibitem{daub_2009}E. G. Daub and J. M. Carlson, Phys. Rev. E \textbf{80}, 066113 (2009).

\bibitem{einav_2007a}I. Einav, J. Mech. Phys. Solids \textbf{55}, 1274.

\bibitem{einav_2007b}I. Einav, J. Mech. Phys. Solids \textbf{55}, 1298.

\bibitem{einav_2009}G. D. Nguyen and I. Einav, Pure. Appl. Geophys. \textbf{166}, 1693 (2009).

\bibitem{lyakhovsky_2013}V. Lyakhovsky and Y. Ben-Zion, J. Mech. Phys. Solids, submitted (2013).

\bibitem{jop_2006}P. Jop, Y. Forterre, and O. Pouliquen, Nature (London) \textbf{441}, 727 (2006).

\bibitem{brodsky_2012}N. J. van der Elst, E. E. Brodsky, P.-Y. Le Bas, and P. A. Johnson, J. Geophys. Res. \textbf{117}, B09314 (2012).

\bibitem{johnson_2008}P. A. Johnson, H. Savage, M. Knuth, J. Gomberg, and C. Marone, Nature \textbf{451}, 57 (2008).

\bibitem{daniels_2005}K. E. Daniels and R. P. Behringer, Phys. Rev. Lett. \textbf{94}, 168001 (2005).

\bibitem{bazant_1999}Z. P. Bazant, Arch. Appl. Mech. \textbf{69}, 703 (1999).

\bibitem{coleman_1963}B. D. Coleman and W. Noll, Arch. Ration. Mech. Anal. \textbf{13}, 167 (1963).

\bibitem{langer_2009a}E. Bouchbinder and J. S. Langer, Phys. Rev. E \textbf{80}, 031131 (2009).

\bibitem{langer_2009b}E. Bouchbinder and J. S. Langer, Phys. Rev. E \textbf{80}, 031132 (2009).

\bibitem{langer_2009c}E. Bouchbinder and J. S. Langer, Phys. Rev. E \textbf{80}, 031133 (2009).

\bibitem{mandl_1977}G. Mandl, L. N. J. de Jong, and A. Maltha, Rock Mech. \textbf{9}, 95 (1977).

\bibitem{pechenik_2003}J. S. Langer and L. Pechenik, Phys. Rev. E \textbf{68}, 061507 (2003).

\bibitem{haxton_2007}T. K. Haxton and A. J. Liu, Phys. Rev. Lett. \textbf{99}, 195701 (2007).

\bibitem{manning_2007b}J. S. Langer and M. L. Manning, Phys. Rev. E \textbf{76}, 056107 (2007).

\bibitem{makse_2004}H. A. Makse, N. Gland, D. L. Johnson, and L. Schwartz, Phys. Rev. E \textbf{70}, 061302 (2004).

\bibitem{walton_1987}K. Walton, J. Mech. Phys. Solids \textbf{35}, 213 (1987).





\end{thebibliography}
\end{document}